# A Brief Survey and an Application of Semantic Image Segmentation for Autonomous Driving

Çağrı Kaymak and Ayşegül Uçar

Firat University, Mechatronics Eng. Dept. 23119, Elazig, Turkey
ckaymak@firat.edu.tr, agulucar@firat.edu.tr

**Abstract.** Deep learning is a fast-growing machine learning approach to perceive and understand large amounts of data. In this paper, general information about the deep learning approach which is attracted much attention in the field of machine learning is given in recent years and an application about semantic image segmentation is carried out in order to help autonomous driving of autonomous vehicles. This application is implemented with Fully Convolutional Network (FCN) architectures obtained by modifying the Convolutional Neural Network (CNN) architectures based on deep learning. Experimental studies for the application are utilized 4 different FCN architectures named FCN-AlexNet, FCN-8s, FCN-16s and FCN-32s. For the experimental studies, FCNs are first trained separately and validation accuracies of these trained network models on the used dataset is compared. In addition, image segmentation inferences are visualized to take account of how precisely FCN architectures can segment objects.

## 1 Introduction

With advanced technology, modern camera systems can be placed in many places, from mobile phones to surveillance systems and autonomous vehicles, to obtain very high quality images at low cost [1]. This increases the demand for systems that can interpret and understand these images.

The interpretation of images has been approached in various ways for years. However, the process involving reviewing images to identify objects and assess their importance is the same [2]. Learning problems from visual information are generally separated into three categories called as image classification [3], object localization and detection [4], and semantic segmentation [5].

Semantic image segmentation is the process of mapping and classifying the natural world for many critical applications such as especially autonomous driving, robotic navigation, localization, and scene understanding. Semantic segmentation, which is a pixel-level labeling for image classification, is an important technique for the scene understanding. Because each pixel is labeled as belonging to a given semantic class.

A typical urban scene consists of classes such as street lamp, traffic light, car, pedestrian, barrier and sidewalk.

Autonomous driving will be one of the revolutionary technologies in the near future in terms of the impact on the lives of people living in industrially developed countries [6]. Many research communities have contributed to the development of autonomous driving systems thanks to rapidly the increasing performance of vision-based algorithms such as object detection, road segmentation and recognition of traffic signals. An autonomous vehicle must sense its surroundings and act safely to reach a certain target. Such functionality is carried out by using several types of classifiers.

Approximately up to the end of 2010, the identification of a visual phenomena was constructed as a two-stage problem. The first of these stages is to extract features from the image. Extensive efforts have been made to extract the features as visual descriptors and consequently the descriptors obtained by algorithms such as Scale Invariant Feature Transform (SIFT) [7], Local Binary Patterns (LBP) [8] and Histogram of Oriented Gradients (HOG) [9] have become widely accepted. The second stage includes to use or design classifier. Artificial Neural Networks (ANNs) are one of the most important classifiers. ANNs are not a new approach and its past is based on about 60 years ago. Until the 1990s, ANNs used in various fields did not provide satisfactory achievements on nonlinear systems. Therefore, there are not many studies about ANNs for a certain period. In 2006, Hinton et al. [10] used ANNs in speech recognition problems and achieved successful results. Thus, ANNs have come up again in the scientific world. Henceforth, researchers thought that the ANNs would be the solution to problems in most areas, but they soon realized that it was a wrong idea with various reasons, such as failure in the training of multi-layer ANNs. Then, the researchers turned to new approaches finding the most accurate class boundaries in feature space and input space such as Support Vector Machine (SVM) [11], AdaBoost [12], and Spherical and Elliptical classifiers [13] using the features obtained from the first stage. In addition to over-detailed class models to facilitate the search for completely accurate boundaries, methods of transforming feature space such as Principal Component Analysis (PCA) and kernel mapping have also been developed.

Later, in image recognition competitions such as the ImageNet Large Scale Visual Recognition Competition (ILSVRC), ANN-based systems took the lead and began to get first place every year by making a big difference to other systems. As time progressed, especially through the internet, very large amount of data has begun to be produced and stored in the digital environment. When processing this huge amount of data, Central Processing Units (CPUs) on the computers have been slow. Along with the developments in GPU technology, the computational operations can be performed much faster by using the parallel computing architecture of the graphics processor. With this increase in process power, the use of deeper neural networks has become widespread in practice. By means of this, “Deep Learning” term has emerged as a new approach in the machine learning.

Deep learning is the whole of the methods consisting of ANNs, which has a deep architecture with an increased number of hidden layers. At each layer of this deep architecture, features belonging to the problem is learned and this learned features create an input into an upper layer. This creates a structure from the bottom layer to the top layer, where the features are learned from the simplest to the most complex. It would be useful to analyze the vision system in the human brain to understand this structure. The signals coming to the eyes through nerves are evaluated in a multi-layer hierarchical structure. At the first layer where the signal is coming after the eyes, the local and basic features of the image, such as the edge and corner, are determined. By combining these features, at the next layer, mouth, nose, etc. details and at the subsequent layers, features belonging to the overall of image, such as face, person and location of objects, respectively can also be determined. Convolutional Neural Networks (CNNs) approach, which combines both feature extraction and classification capabilities in computer vision applications, work in this way.

Deep learning brings the success of artificial intelligence applications developed in recent years to very high levels. Deep learning is used in many areas such as computer vision, speech recognition, natural language processing and embedded systems. In the ILSVRC, which has been carried out using huge data sets in recent years, the competitors have been directed to the CNN approaches and achieved great success [14]. Companies such as Google [15], Facebook [16], Microsoft [17] and Baidu [18] have realized the progress in deep learning and carried out studies on this topic with great investments.

A graphical representation of search interest of the "Deep Learning" on the Google search engine in the last 5 years is shown in Fig. 1.

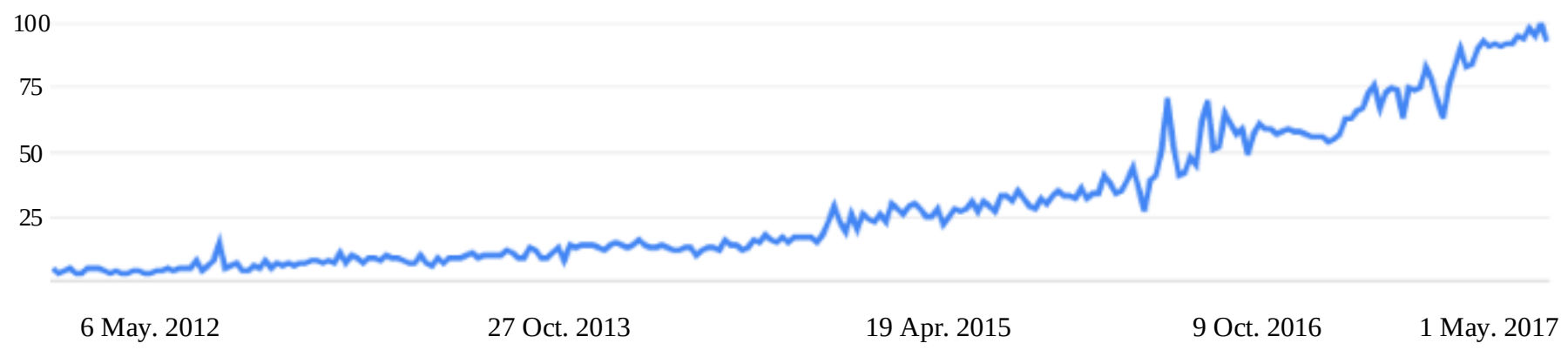

**Fig. 1.** Search interest of the "Deep Learning" on the Google search engine in the last 5 years [19]

The advancement of CNNs is based on a high amount of labeled data. In general terms, CNNs carry out end-to-end learning by predicting class labels from raw image data by learning millions of parameters, which is more successful than methods based on visual descriptors.

In semantic image segmentation, a large number of studies have recently been conducted [20-32] to overcome of the supervised semantic segmentation using images with pixel-pixel annotations to train the CNNs. Some of the semantic segmentation studies was tried to directly adopt CNN architectures designed for image classification. However, the results were not very satisfactory. Because standard CNN architectures are not suitable to the semantic segmentation due to loss of the spatial position. While on one hand, repeated convolutional and pooling layers reduce the spatial resolution of feature maps, on the other hand, fully connected layers produce class probability values by completely discarding spatial information to produce an output.

In recent years, a lot of network architectures have emerged that are capable of bringing semantic information to a pixel location [20-21]. First, Long, Shelhamer and Darrell [20] converted the pre-trained VGG-16 [33] CNN architecture for classification into Fully Convolutional Networks (FCNs). For this, they replaced all fully connected layers of the CNN architecture with convolutional layers and added deconvo-lutional layers that restore the original spatial resolution with the skip connections. Thus, three different versions of FCN called FCN-8s, FCN-16s and FCN-32s were obtained, and the spatial resolution at the output was brought to the spatial resolution of the input image to generate the class probability values for each pixel.

The lack of a deep deconvolutional network trained in a large dataset makes it difficult to completely reconstruct nonlinear structures of object boundaries. Chen et al. [21] have contributed to correct this problem by applying the Conditional Random Field (CRF) method to the output of FCN.

Noh, Hong and Han [22] constructed FCN architecture named DeconvNet by using the convolutional layers adopted from the VGG-16 network architecture with the proposed deconvolutional network architecture. DeconvNet performed well on the PASCAL VOC 2012 [21] dataset.

Badrinarayanan, Kendall and Cipolla [27] proposed a new and practical FCN architecture called SegNet for semantic segmentation. This architecture consists of the encoder, which is the trainable segmentation unit, followed by the corresponding decoder and classifier layer. In the encoder network architecture, 13 convolutional layers in the VGG-16 network architecture were used likewise. They compared the SegNet architecture with the original FCNs [20] and DeconvNet [22] architectures on the SUN RGB-D [34] and CamVid [35] road scene datasets. They provided that the SegNet architecture has fewer trainable parameters than other FCN architectures and is therefore efficient both in terms of memory and computational time. In [28], they implemented the Bayesian SegNet architecture, an extension of SegNet, on the same datasets. The architecture resulted in improving of the boundary lines, increasing the accuracy of prediction, and reducing the number of parameters.

Fourure et al. [29] presented an approach that is enhanced by multiple datasets to improve the semantic segmentation accuracy on the KITTI [36] dataset used for autonomous driving. To take advantage of training data from multiple datasets with different tasks including different label sets, they proposed a new selective loss function that can be integrated into deep networks.

Treml et al. [30] conducted a study to reduce the computational load of embedded systems found in autonomous vehicles for autonomous driving. They designed a network architecture that preserves the accuracy of semantic segmentation while reducing the computational load. This architecture consists of an encoder like SqueezeNet [37], Exponential Linear Unit (ELU) used instead of Rectified Linear Unit (ReLU) activation function and a decoder like SharpMask [38].

Hoffman et al. [31] trained three different FCN architectures in [20] on both GTA5 [39] and SYNTHIA [40] datasets to examine adaptations simulated real images in CityScapes [41] and compared inference performances.

Marmanis et al. [42] proposed a FCN architecture for the semantic segmentation of very high resolution aerial images. They initiated the FCN with learned weight and bias parameters using FCN-PASCAL [20] network model pre-trained on the PASCAL VOC 2012 dataset.

The rest of this paper is organized as follows. In Section 2, the necessary concepts for a better understanding of deep learning are introduced in detail. In Section 3, firstly, the structures that constitute the CNN architectures are explained, next, information about the training of the CNN and the necessary parameters affecting its performance are given. In Section 4, it is explained how to make the conversion from CNN to FCN used in paper by explaining the main differences between image classification and semantic image segmentation applications. The Section 5 gives information about the dataset used for the semantic image segmentation application and the experimental results obtained by FCN architectures. In Section 6, the paper is concluded.

## 2 Deep Learning

Deep learning is a fast-growing popular machine learning approach in the artificial intelligence field to create a model for perceiving and understanding large quantities of machines, such as images and sound. Basically, this approach is based on deep architectures, which are the more structurally complex of the ANNs. This deep architecture term refers to ANNs whose number of hidden layers has been increased.

Deep learning algorithms are separated from existing algorithms in machine learning; it needs very high amount of data and hardware with very high computational power that can handle this high data rate. In recent years, the number of labeled images, especially in the field of computer vision, has increased extremely. Deep learning approach has attracted much attention thanks to the great progress in the area of GPU-based parallel computing power. GPUs with thousands of compute cores provide 10 to 100 times the application performance when processing these data compared to CPUs [43]. Nowadays, deep learning has many application areas, mainly automatic speech recognition, image recognition and natural language processing.

There are many different types of deep learning architecture. Basically, deep learning architectures can be named as in Fig. 2 [44].

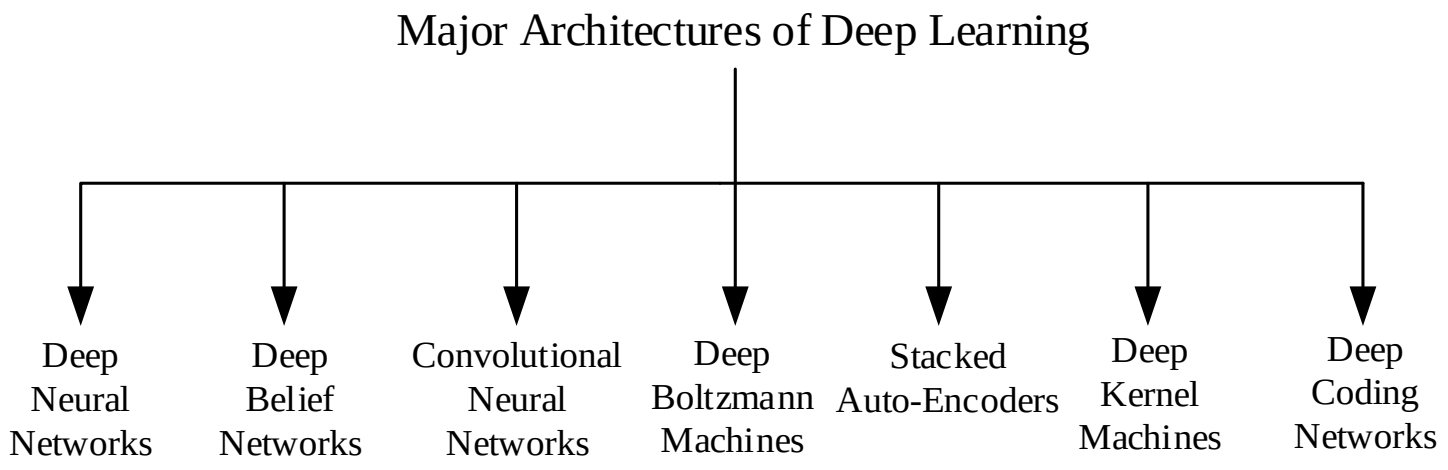

**Fig. 2.** Major architectures of deep learning

In order to be understand better the deep learning term, it is necessary to adopt ANN structures in a good way. For this reason, this information will be given first. In addition, we will focus on the feedforward ANN because the FCNs are the multi-layer feedforward neural network type from the deep learning architectures that form the basis of our paper.

### 2.1 Artificial Neural Networks

ANNs have been developed in the light of the learning process in the human brain. As the neurons in the biological nervous system connect with each other, the structures defined as artificial neurons in the ANN systems are modeled to be related to each other. ANNs can be used in many areas such as system identification, image and speech recognition, prediction and estimation, failure analysis, medicine, communication, traffic, production management and more.

#### 2.1.1 Neuron

ANNs also have artificial neurons, as biological neural networks are neurons. The neuron in can be called the basic calculation unit in ANN. The neurons can also called node or unit. The structure of an artificial neuron is shown in Fig. 3.

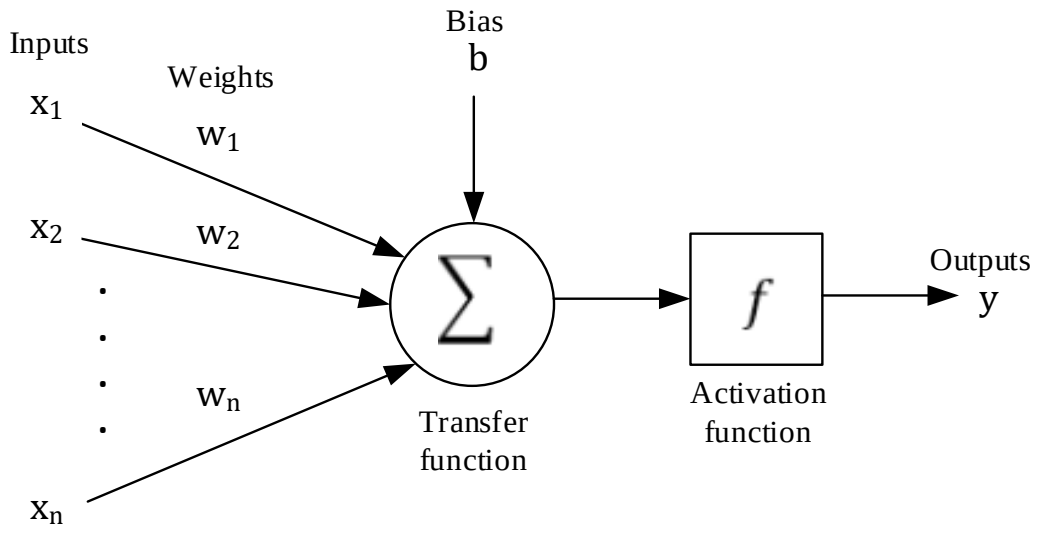

**Fig. 3.** The structure of an artificial neuron

Inputs are information incoming to a neuron from external world. These are determined by the samples for which the learning of the network is desired. Weights show the importance of information incoming to neurons and their effect on neurons. There are separate weights for each input. For example, $W_1$ weight in Fig. 3 shows the effect of $x_1$ input on the neuron. The fact that weights are big or small does not mean that they are important or insignificant. The transfer function calculates the net input incoming to a neuron. Although there are a large number of transfer functions for this, the most commonly used is weighted sum. Each incoming information is summed by multiplying its own weight. The activation function determines the output the neuron will generate in response to this input by processing the net input incoming to the neuron. The generated output is sent to the external world or another neuron. In addition, if desired, the neuron may also send its own output as an input to itself.

The activation function is usually chosen a nonlinear function. The purpose of the activation function is to transfer the nonlinearity to the output of the neuron as in (1). A characteristic of ANNs is nonlinearity, which is due to the nonlinearity of activation functions.

$$y = f\left(\sum_{i=1}^{n} W_i x_i + b\right) \tag{1}$$

The important thing to note when choosing the activation function is that the derivative of the function is easy to calculate. This ensures that the calculations take place quickly.

In the literature, there are many activation functions such as linear, step, sigmoid, hyperbolic tangent (tanh), Rectified Linear Unit (ReLU) and threshold functions.

However, sigmoid, tanh and ReLU activation functions are usually used in ANN applications.

The sigmoid activation function, expressed by (2), is a continuous and derivatable function. It is one of the most used activation functions in ANNs. This function generates a value between 0 and 1 for each input value.

$$\sigma(x) = \frac{e^x}{1 + e^x} \tag{2}$$

The tanh activation function, expressed by (3), is similar to the sigmoid activation function. However, the output values range from -1 to 1.

$$\tanh(x) = 2\sigma(2x) - 1 \tag{3}$$

The ReLU activation function, expressed by (4), generates an output with a threshold value of 0 for each of the input values. It has a characteristic as in Fig. 4. Recently, the usage in ANNs has become very popular.

$$f(x) = \max(0, x) \tag{4}$$

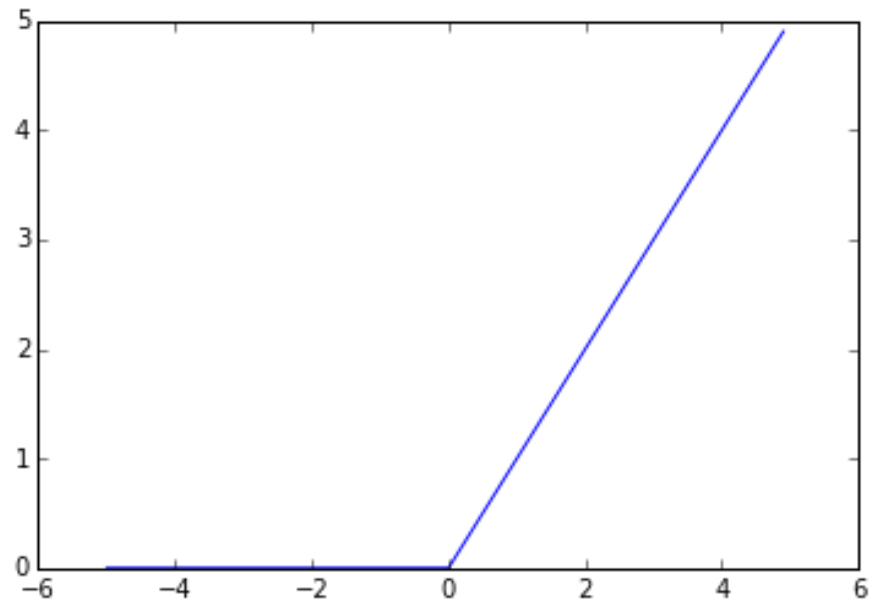

**Fig. 4.** The characteristic of ReLU activation function

### 2.1.2 Feedforward Neural Network

In the feed forward neural network, the flow of information is only in the forward direction. The neurons in the network are arranged in the form of layers and the outputs of the neurons in a layer are input to the next layer via weights. The feedforward neural networks are basically composed of 3 types of layers. These layers are input, hidden and output layer. The input layer transmits the information incoming from the external world to the neurons in the hidden layer without making any changes. This information is then sequentially processed in the hidden layer/layers, which are not associated with the external world, and the output layer, which transfers the information from the network to the external world, to determine the network output in response to the desired input.

A feedforward neural network may have one or more hidden layers or no hidden layers. If the network does not contain any hidden layers, it is called single-layer perceptron, if it contains one or more hidden layers, it is called multi-layer perceptron.

In Fig. 5, a 3-layer feedforward ANN model is shown as an example.

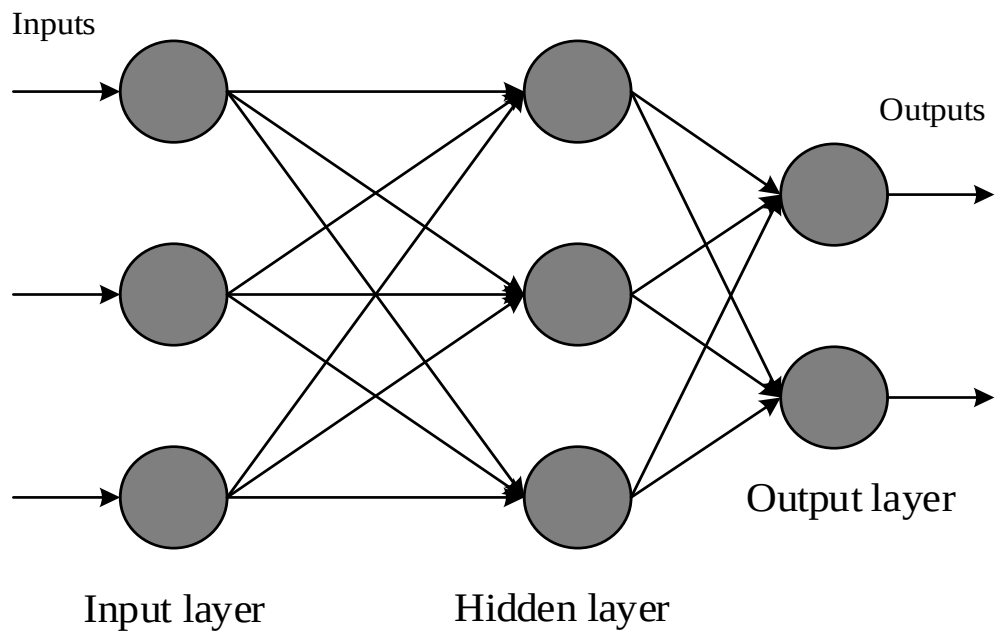

**Fig. 5.** 3-layer feedforward ANN model

In a multi-layer feedforward neural network, each neuron is only associated with the next neurons. In other words, there is no connection between the neurons in the same layer.

The term of depth in a multi-layer feedforward neural network is related to the number of hidden layers. As the number of hidden layers of the network increases, the depth increases. In short, a network with multiple hidden layers can be expressed as a deep neural network. Fig. 6. shows a deep feedforward ANN model as an example.

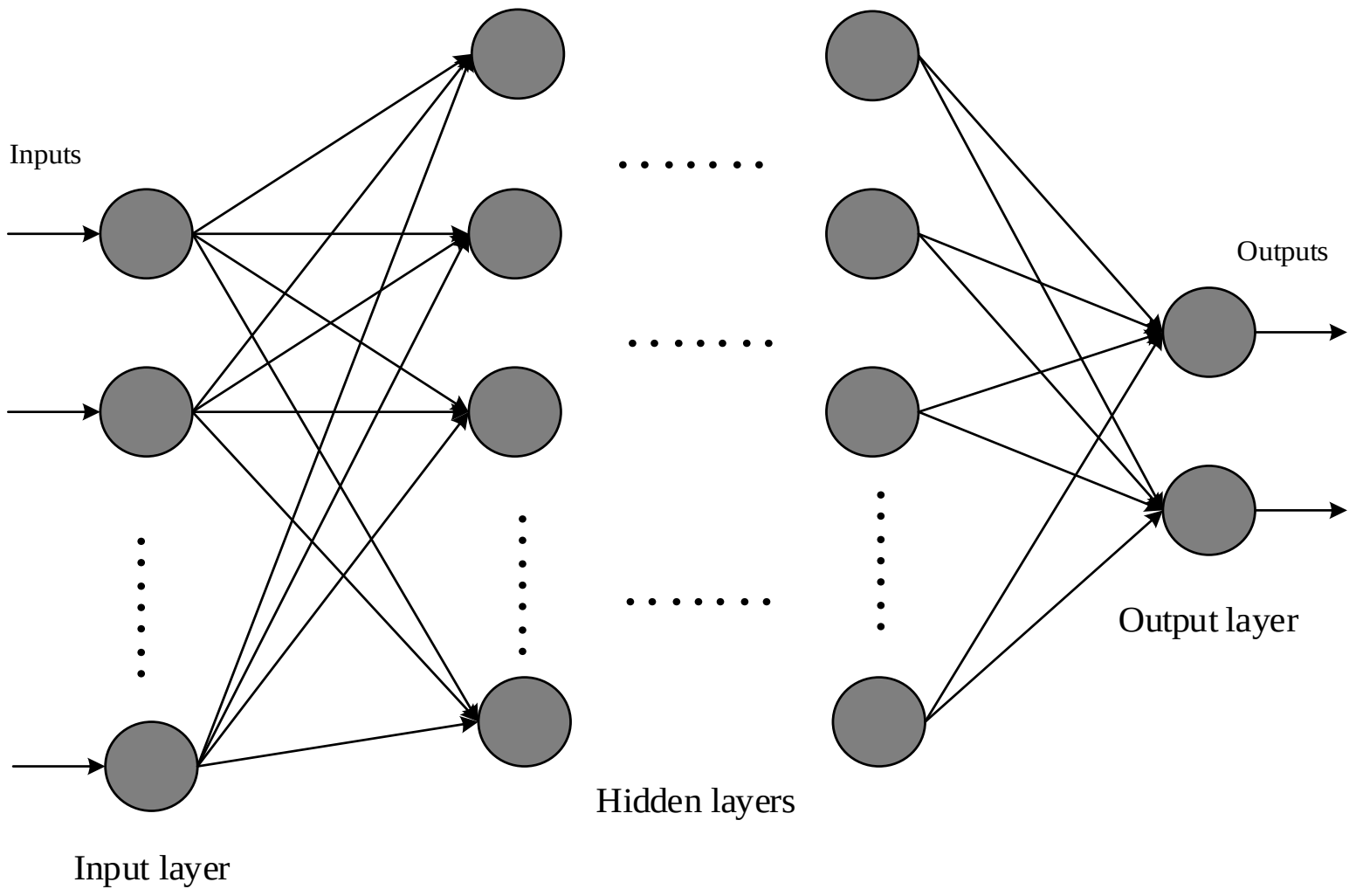

**Fig. 6.** A deep feedforward ANN model

The ability to learn from an information source is one of the most important features of ANN. In multi-layer neural networks, learning process takes place by changing weights at each step. Therefore, how weights are determined is important. Since the information is stored in the entire network, the weight value of a neuron does not make sense by itself. The weights on the whole network should get the most appropriate values. The process to achieve these weights is to train the network. In short, the learning of the network occurs by finding the most appropriate values of the weights. In addition, there are a number of considerations to be taken when designing multi-layer neural networks, such as the number of hidden layers in a network, the number of neurons to be found in each hidden layer, the optimal solution for the most reasonable time, and the test of network accuracy [45].

## 2.2 Deep Learning Software Frameworks and Libraries

The workflow of deep learning is a multi-stage, iterative process. In this process, the data must first be collected and preprocessed if necessary. Large-scale datasets may be needed for to take place a successful deep learning. Nowadays, thanks to the internet and large data sources, datasets have been growing rapidly. The network is trained using these datasets. Existing networks can be used to train the network, or new networks can be developed. The network models created after the training phase must be tested to confirm that they work as expected. Generally at this point, it is necessary to repeat certain operations to improve the results. These operations include reprocessing the data, arranging the networks, changing the parameters of the networks or solvers, and retesting until the desired result is obtained. The CPUs of these computers are insufficient for these intensive computational processes to be performed in the deep learning process. Because the CPUs with a certain processing capacity and architecture cannot perform many operations at the same time, the training and test phases of the model take a lot of time. Because of this, CPUs have given place to GPUs that allow parallel processing of data. By means of this, deep learning has begun to be used quickly in real life applications.

In the deep learning applications, NVIDIA provides a CUDA extension that allows GPUs to perform parallel computing [46]. CUDA is a parallel computing architecture that uses NVIDIA's GPU power to accelerate computing performance at a high level. CUDA enables the usage of graphics processor cores for general purpose accelerated computing.

There are many popular software frameworks and libraries, especially including Caffe, Torch, Theano, TensorFlow, Keras and DIGITS, for the implementation of deep learning algorithms. Most of them can also run on the GPU.

### 2.2.1 Caffe

The Caffe deep learning framework, created by Yangqing Jia, is developed by the Berkeley AI Research (BAIR) and community contributors. Caffe was designed to be as fast and modular just like the human brain [47].

Caffe is often preferred in industrial and academic research applications. The most important reason for this is the ability to process data quickly. Caffe can process over 60 million images per day with a single NVIDIA K40 GPU. Caffe is believed to be among the fastest accessible CNN implementations available [47].

### 2.2.2 Torch

Written in LuaJIT language, Torch is a scientific computing structure that provides extensive support for machine learning algorithms. It is an easy and efficient library because it is written in LuaJIT and uses the C/CUDA application basis [48]. This library, which can use numerical optimization methods, contains various neural networks and energy based models. It is also open source and provides fast and efficient GPU support.

Torch is constantly being developed and is being used by various companies such as Facebook, Google and Twitter.

### 2.2.3 Theano

Theano is a Python library that effectively identifies, evaluates, and optimizes mathematical expressions containing tensors [49]. Since this library is integrated with NumPy library, it can easily perform intensive mathematical operations. It also offers the option to create dynamic C code, allowing user to evaluate expressions more quickly.

### 2.2.4 TensorFlow

Tensorflow is an open source deep learning library that performs numerical computations using data flow graphs. This library was developed by Google primarily to conduct research on machine learning and deep neural networks [50]. With its flexible architecture, TensorFlow allows you to deploy the computation to one or more CPUs or GPUs on a server, mobile or desktop device with a single Application Programming Interface (API).

Snapchat, Twitter, Google and eBay, which are popular nowadays, also benefit from TensorFlow.

### 2.2.5 Keras

Keras is a modular Python library built on TensorFlow and Theano deep learning libraries [51]. These two basic libraries provide the ability to run on the GPU or CPU. By making minor changes in the configuration file of Keras, it is possible to use the TensorFlow or Theano in the background.

Keras is very useful as it simplifies the interface of TensorFlow and Theano libraries, and easier application can be developed than these two libraries. Keras has a very common usage in image processing applications.

### 2.2.6 DIGITS

In 2015, NVIDIA introduced the CUDA Deep Neural Network library (cuDNN) [52] due to the growing importance of deep neural networks, both in the industrial and academia, and the great role of GPUs. In 2016, Jen-Hsun Huang, NVIDIA CEO and founder, has brought the Deep Learning GPU Training System (DIGITS) into use at the GPU Technology Conference.

DIGITS is a deep learning GPU training system that helps users to develop and test CNNs. This system supports GPU acceleration using cuDNN to greatly reduce training time while visualizing Caffe, Torch and TensorFlow by providing web interface support.

DIGITS supports many educational objectives including image classification, semantic segmentation and object detection. Fig. 7 shows the main console window where datasets can be generated from the images and they can be prepared for training. In DIGITS, once a dataset is available, the network model can be configured and training can begin. DIGITS also provides the necessary tools for network optimization. Settings for network configuration can be followed and accuracy can be maximized by changing parameters such as bias, activation functions and layers.

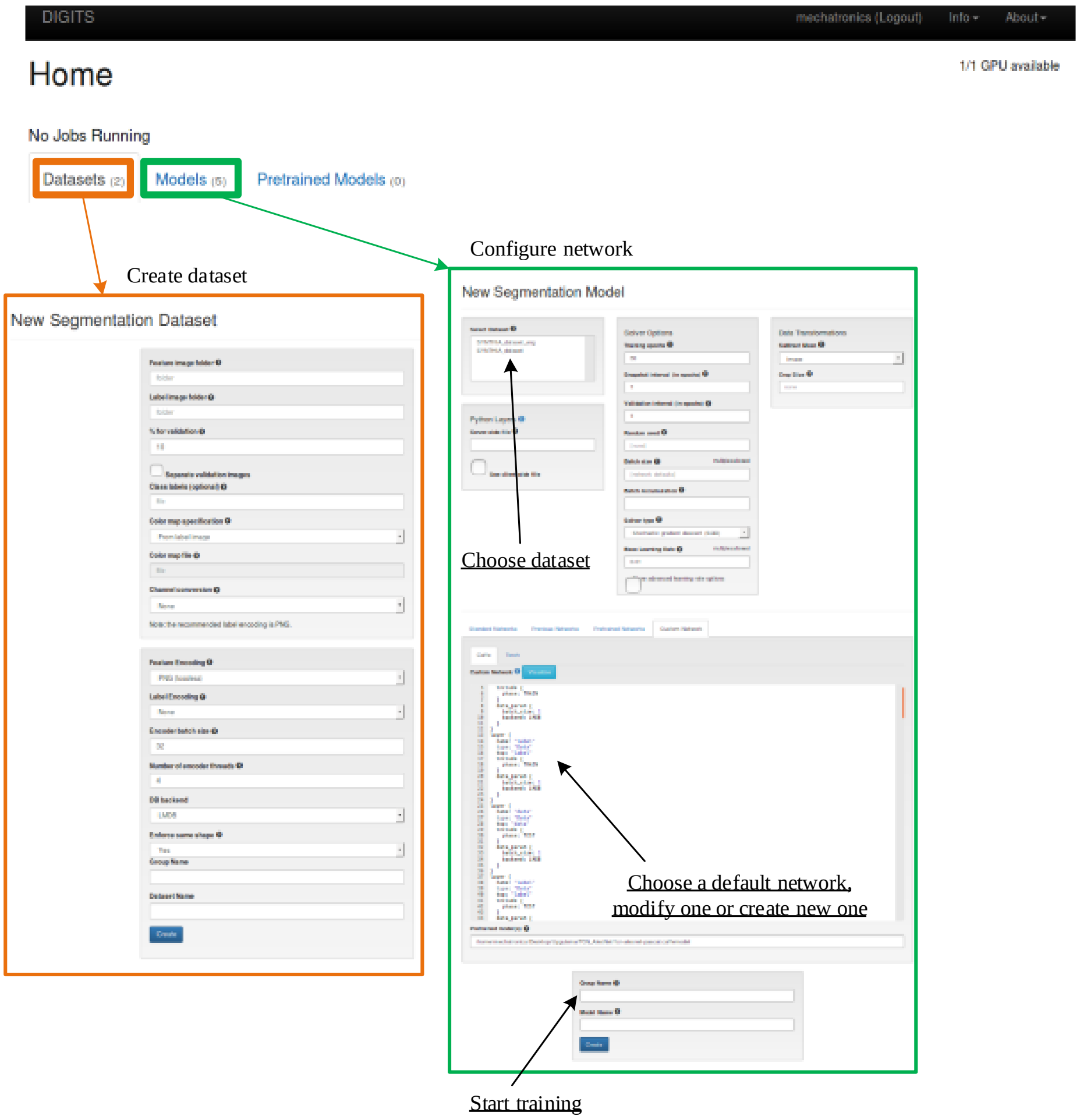

**Fig. 7.** DIGITS main console

## 3 Convolutional Neural Networks

CNNs, introduced by LeCun in 1989 for computer vision applications, are a type of multi-layer feedforward-ANN [53]. Nowadays, CNNs become increasingly popular among in-deep learning methods as they can successfully learn models for many computer and visual applications such as object detection, object recognition, and semantic image segmentation.

CNNs can be thought of as classifiers that extract hierarchical features from raw data. In CNN, images are given as input to the network, and learning takes place automatically with a feature hierarchy created without using any feature extractor method.

### 3.1 Architecture

All neurons in a layer in the feedforward ANNs are connected to all neurons of the next layer. Such connected layers are called fully connected layers and, in addition to fully connected layers in the CNN, convolution is applied to the input image to generate an output. This is caused by the local connection that all regions in the input layer are bound to neurons in the next layer. Thus, the input image is convolved with each learned filter used in this layer to generate different feature maps. The feature maps become more insensitive to rotation and distortion by providing more and more complex generalizations towards higher layers. In addition, the feature maps obtained in the convolutional layer are subjected to the pooling layer in order to perform spatial dimensionality reduction and keeping of important features. A classifier always is the final layer to generate class probability values as an output. The final output from the convolutional and pooling layers is transferred to one or more fully connected layers. Then, the output prediction is obtained by transferring to the classifier layer where the activation functions such as Softmax are used.

A simple CNN architecture is a combination of convolutional, pooling and fully connected layers as in Fig. 8.

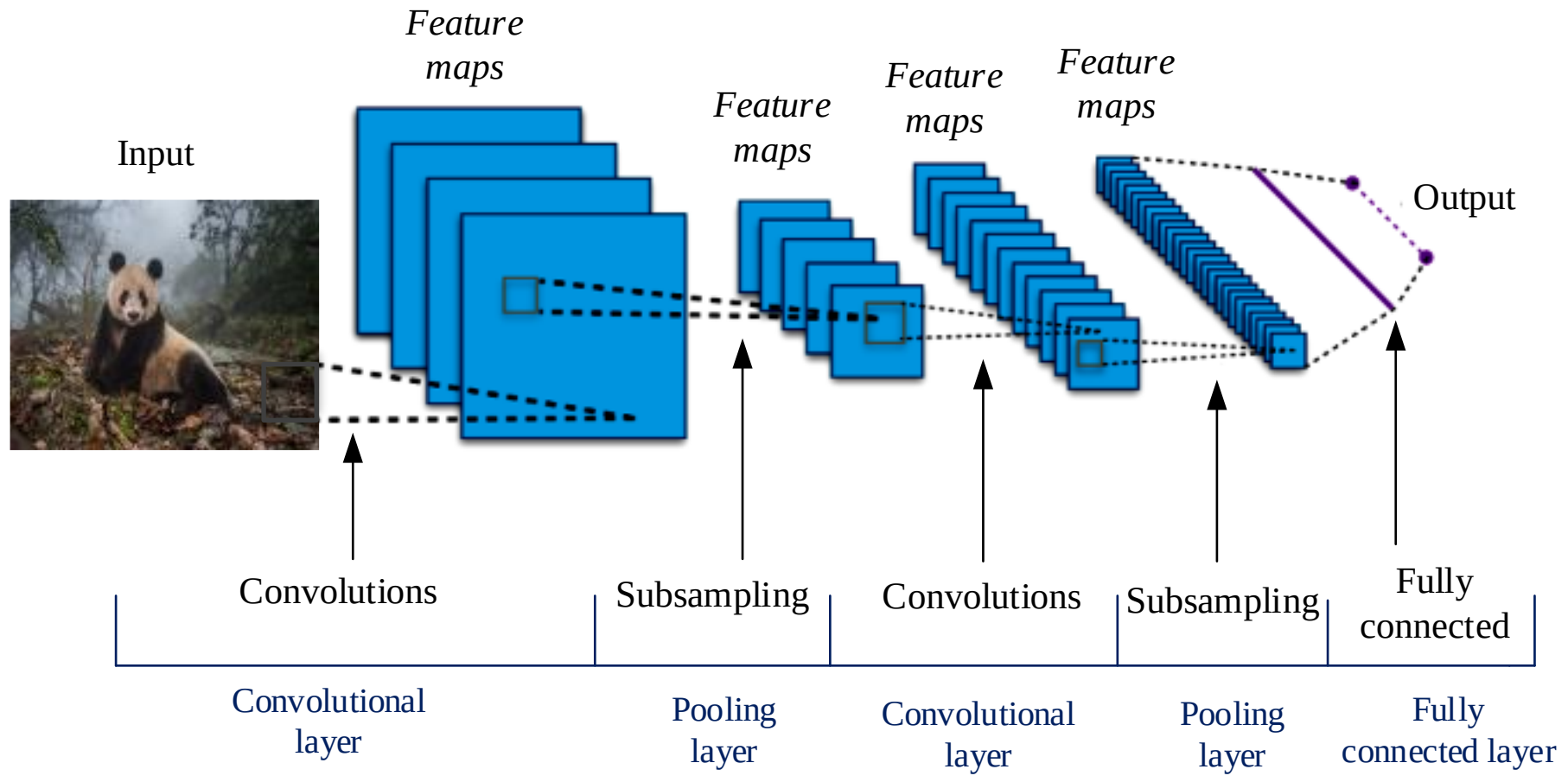

**Fig. 8.** An example of a simple CNN architecture

### 3.1.1 Convolutional Layer

The purpose of the convolutional layer, which is the most basic layer of CNNs, is to convolve the input image with learnable filters and extract its features. A feature map is generated with each filter. CNNs draw attention to the fact that when applied to RGB images (images used in this paper), the image is a 3D matrix, and each of the layers is similarly arranged. This is shown in Fig. 9. Each layer of CNN consists of a set of spatial filters of size d×h×w that are the spatial dimensions of h and w that appear as volume of the neurons and the number of kernel (or filter) feature channels of d. Each of these filters is subjected to convolution with a corresponding volume of the input image, and slid through the entire image (sized $D_i \times H_i \times W_i$ where $H_i$, $W_i$ are the spatial dimensions and $D_i$ is the channel number) across its spatial dimensions $H_i$,$W_i$. Convolution refers to the sum of element by element multiplication of the neurons in each filter with the corresponding values at the input. Thus, it can be assumed that the first layer in the CNNs is the input image. Based on this, convolution with a single filter in each layer provides a 2-dimensional output with parameters such as stride and padding. This is expressed as a feature map or activation map for a filter in input. At each convolutional layer of the CNNs, N filters are used, each resulting in a feature map. These feature maps are stacked together in a certain volume to obtain the output of a convolutional layer.

A single neuron in a filter of a certain layer can be mapped to connected neurons in all previous layers, following such convolutions. This is called the effective receptive field of the neuron. It is easy to see that convolutions result in very local connections with neurons in the lower layers (closer to the input) with those with smaller receptive fields than in the higher layer. While the lower layers learn to represent small areas of the input, the higher layers learn more specific meanings because they respond to a larger subdivision of the input image. Thus, a feature hierarchy is generated from the local to the global.

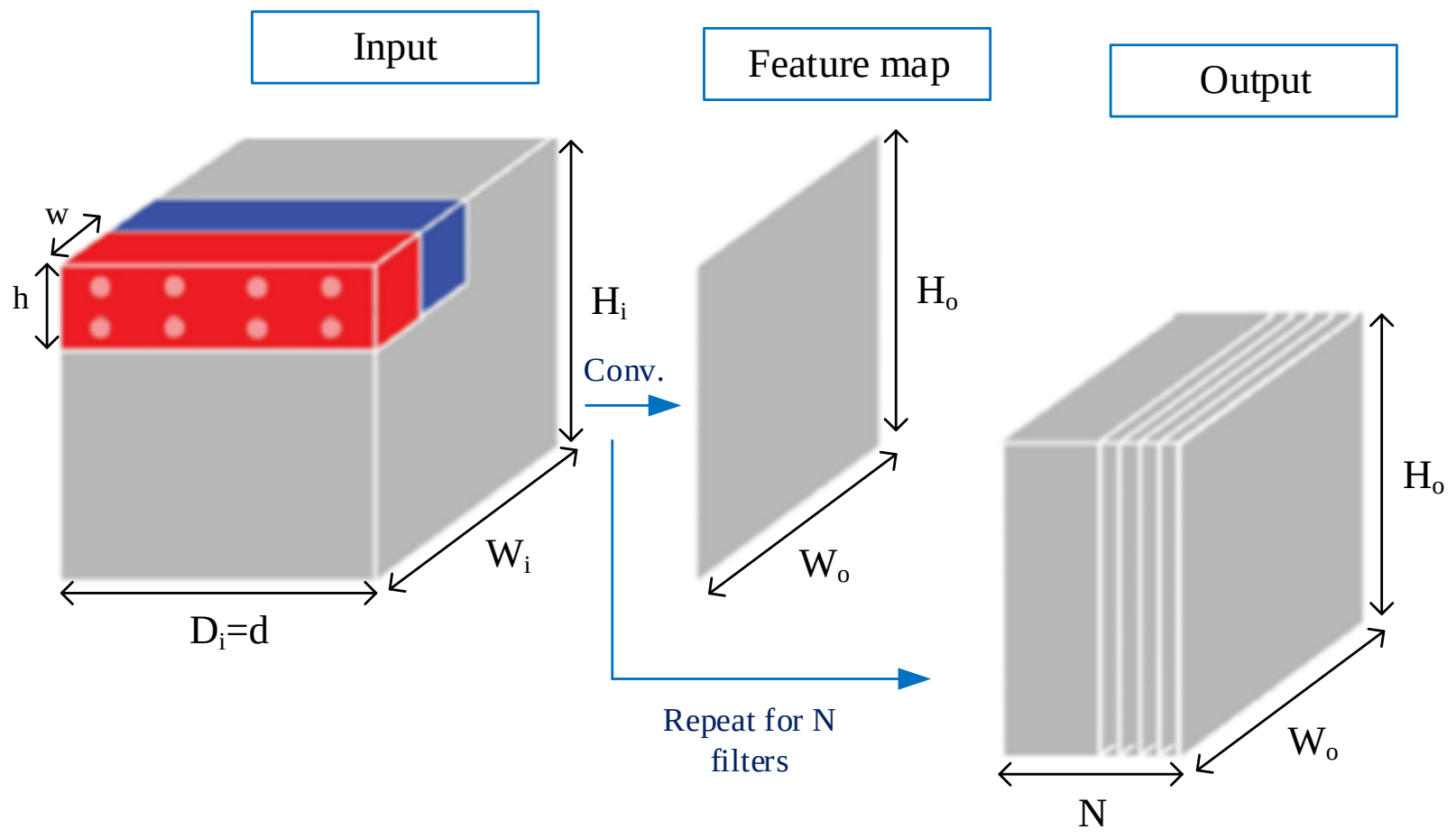

**Fig. 9.** An illustration of a single convolutional layer [54]

The red and blue areas in Fig. 9 represent the two positions of the same filters of size d×h×w that are subjected to convolution by sliding through the input volume. Given that the filter size is 2×2, it can be seen that the stride parameter s is 2. For RGB input image,$D_i$=d=3.

The stride s of a filter is defined as the intervals at which the filter moves in each spatial dimension. p padding corresponds to the number of pixels added to the outer edges of the input. Hence, stride can be considered as an input means of the subsampling [55]. Typically, square filters of the form h=w=f are used. The output volume of such a layer is calculated using equations (5), (6) and (7).

$$D_o = N \tag{5}$$

$$H_o = \frac{H_i - f + 2p}{s} + 1 \tag{6}$$

$$W_o = \frac{W_i - f + 2p}{s} + 1 \tag{7}$$

Fig. 10 shows a 3×3 filter to be slid over a 5×5 image matrix representing a binary image. The sliding of the filter is from left to right and continues until the end of the matrix. In this paper, the stride is taken as 1. By sliding filters in order, the process is completed and the final state of the feature map is obtained as shown in Fig. 10.

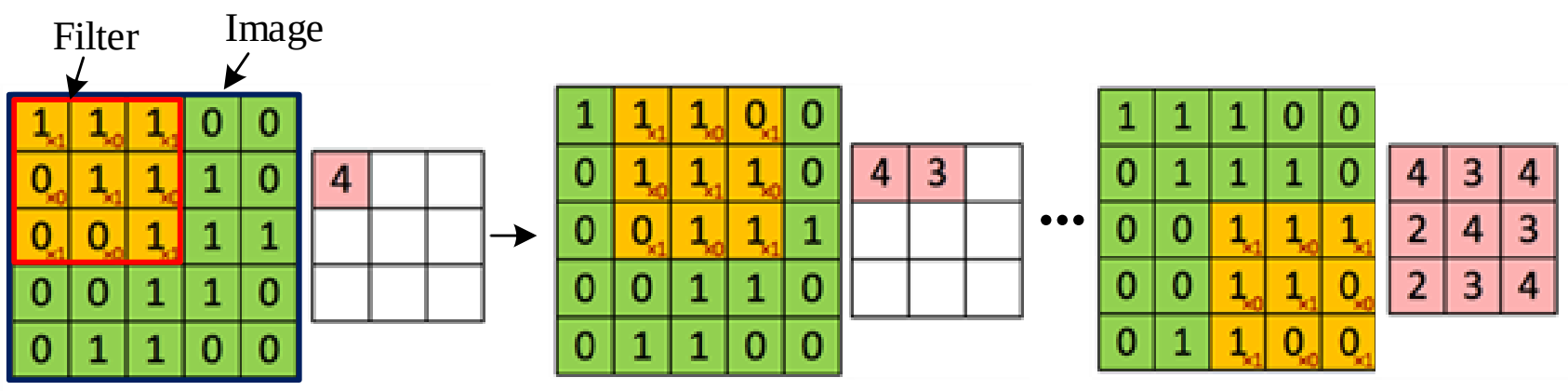

**Fig. 10.** Image matrix and final state of the feature map [56]

### 3.1.2 Pooling Layer

Frequently between the convolutional layers, pooling layers are scattered which help to spatially subsample the input features. The pooling layers make the subsampling process of the input image. This is done by sliding a filter over the input image.

The input image (usually non-overlapping) is divided into subregions and each subregion is sampled by non-linear pooling functions. The best-known of these functions are the maximum and average pooling functions. With the maximum pooling function used throughout this paper, the maximum value is returned from each subregion. The avarage pooling function returns the average value of the subregion. The pooling provides robustness to the network by reducing the amount of translational variance in the image [3]. In addition, unnecessary and redundant features are also discarded, which reduces the network’s computational cost and, therefore, makes it more efficient.

The pooling layers also have a stride parameter that provides control over the output sizes. The same equations used for the output size of the convolutional layers can be used for this layer. It can be seen in Fig. 11 that the input volume of 64×224×224 is subsampled to the volume of 64×112×112 by 2×2 filters and strides 2. The pooling operation is performed separately for each feature map, and the size of the feature map is reduced as shown.

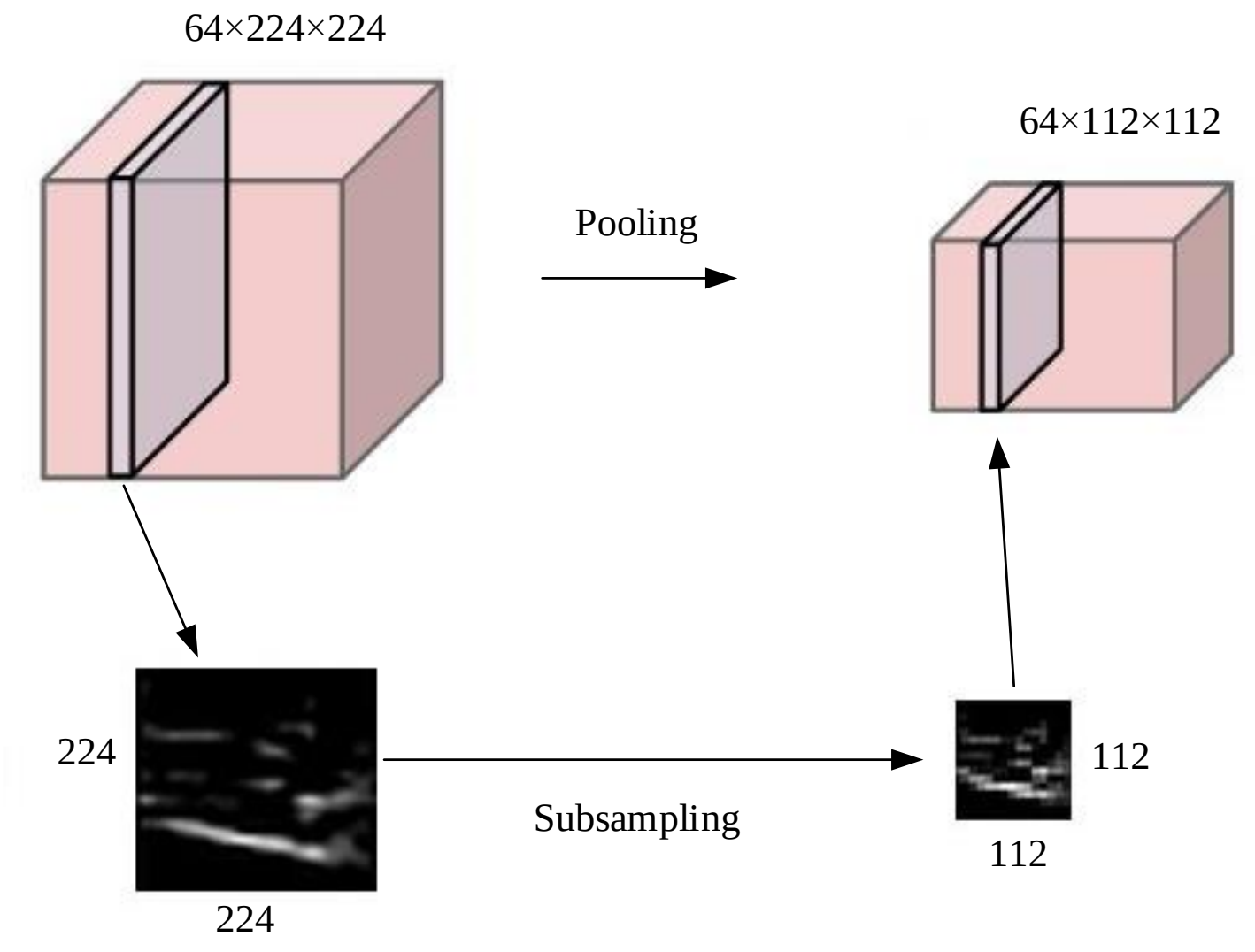

**Fig. 11.** An example of a subsampling [5]

Fig. 12 shows the pooling operation performed with the maximum pooling function by sliding the 2×2 filter, stride 2.

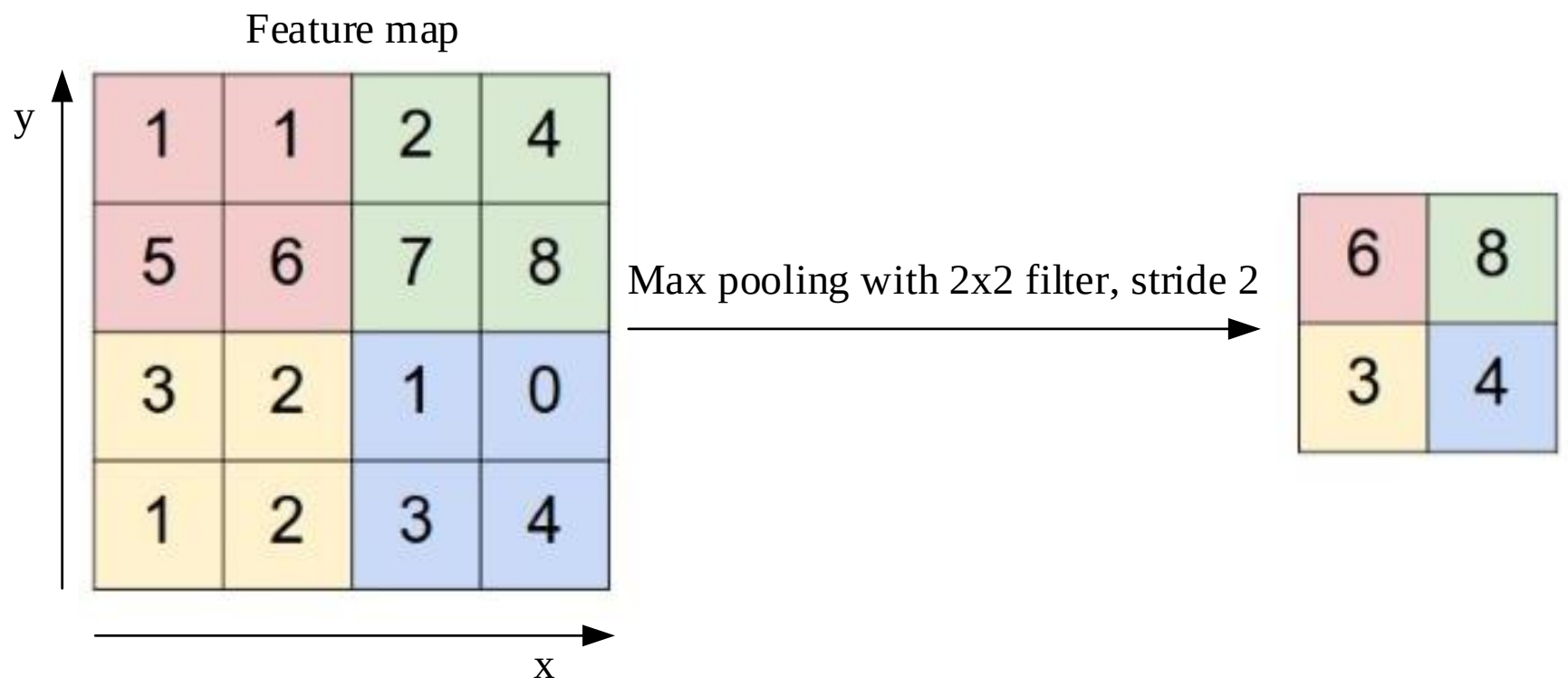

**Fig. 12.** The pooling operation with the 2×2 filter, stride 2 [5]

### 3.1.3 Rectified Linear Unit Layer

Generally, the outputs of the convolutional layer are fed into activation functions. The nonlinearity layers proposed for this purpose can be composed of functions such as sigmoid, tanh and ReLU. ReLU has been found to be more effective in CNNs and is often preferred [57].

A ReLU layer thresholds negative inputs to 0 and activates the positive inputs as described (8) by passing them unchanged.

$$f(x) = \begin{cases} x, & x \geq 0 \\ 0, & \text{others} \end{cases} \tag{8}$$

where; x is the input of the ReLU, and f (x) is the rectified output.

In the ReLU layer, an operation is performed separately for each pixel value. For example, the output of the ReLU is as shown in Fig. 13 if it is considered that the black areas are represented by negative pixels and the white areas are represented by positive pixels in the input feature map.

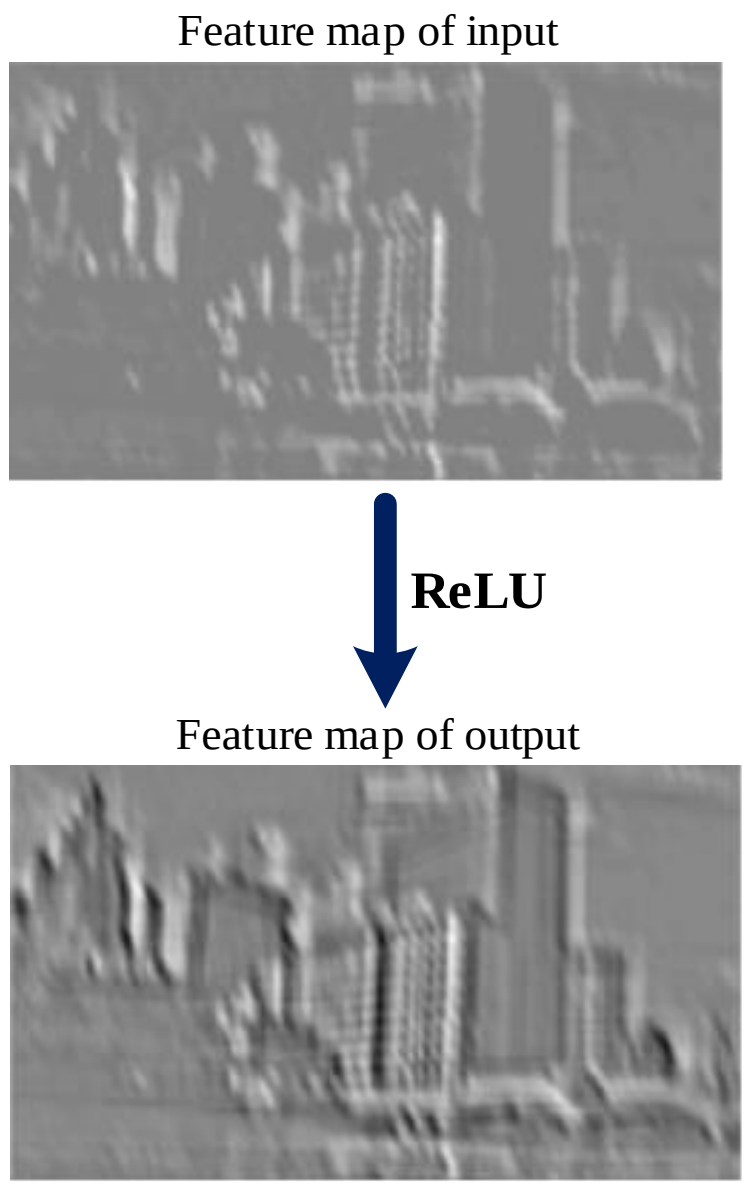

**Fig. 13.** An example of ReLU operation [58]

### 3.1.4 Fully Connected Layer

After high-level features are extracted with convolutional, pooling and ReLU layers, generally the fully connected layer is placed at the end of the network. The neurons in this layer are completely dependent on all activations in the previous layer. The most important feature of the fully connected layer is that it allows the neurons in this layer to determine which features correspond to which classifications. In short, the fully connected layer can be thought of as the layer that feeds the classifier.

Spatial information is lost as a neuron in the fully connected layer receives activations from all input neurons. This is not desirable in this semantic image segmentation paper where spatial information is very important. One way to get over this situation is to see the fully connected layer as a corresponding convolutional layer. It is also based on the basis of the FCNs that will be mentioned in the following.

#### 3.1.5 Classifier

A classifier is chosen by considering the problem at hand and the data used. In this paper, the Softmax function is used which allows to predict for a class other than exclusive classes mutually. For the binary class problem, the Softmax function is reduced to a logistic regression. The Softmax function gives the probability value in (9) for a certain input belonging to a certain class c.

$$p_c = \frac{e^{(s_c)}}{\sum_{i=1}^{c} e^{(s_i)}} \tag{9}$$

where; s is the network outputs obtained from previous layers of CNNs for a particular class. For a single input, the sum of all probabilities between classes is always equal to 1. The loss metric is defined as the probability of the negative logarithm of the Softmax function. This is a cross entropy loss.

#### 3.1.6 Regularization

Overfitting over training data is a big issue. Especially when dealing with deep neural networks that the network is strong enough to fit the training set alone is a big problem. The overfitting must be avoided. The methods developed for this are called regularization methods.

Dropout is a simple and effective regularization strategy integrate into the training phase. Dropout, introduced for the first time in [59], is implemented as dropout layers characterized by a probability value. Dropout can be accepted as a reasonable default value of 0.5 proven to be sufficiently effective [59].

### 3.2 Training

The learning process in CNNs can be divided into 4 basic steps:

1. Forward computation,
2. Error/loss optimization,
3. Backpropagation,
4. Parameter updates.

Forward computation is usually the case where sublayers composed of convolutional or pooling layers are followed by higher fully connected layers. The network returns the class output, which encodes the probability of belonging to a particular class for input. The outputs of the network may be unscaled as in the SVM classifier, or a negative logarithm probability may be obtained as in the Softmax classifier. For semantic segmentation, each pixel in the image is provided with a class output.

The set of class outputs provided by the network should be subjected to optimization processing by adjusting the values of the learned parameters such as weight filters and biases. The uncertainty that occurs in determining which set of parameters is ideal is quantified by the loss function that can be formulated as an optimization problem. For each vector of the class outputs s, the cross entropy loss is calculated as given in (10) for the Softmax classifier.

$$L_i = -\log\left(\frac{e^{(s_c)}}{\sum_{i=1}^{c} e^{(s_i)}}\right) \tag{10}$$

The evaluation function of the cross entropy is calculated by (11).

$$H(p, q) = -\sum_{x} p(x)\log q(x) \tag{11}$$

where; q is the Softmax function defined in (9), and p is the probability distribution. The total loss is calculated by (12).

$$L = \sum_{i=1}^{N} L_i + \lambda R(W) \tag{12}$$

where; N is the number of training samples, λ is the regularization strength, L is the total loss, and R(W) is the regularization term. To minimize L, the problem is formulated as an optimization step and the loss function is minimized. Thus, the probability is maximized.

Backpropagation is a fundamental concept in learning with neural networks. The purpose of backpropagation is to periodically update the initial weight parameters. The backpropagation of the problem helps to optimize the cost function. The optimization algorithm is generally used to understand the gradient descent and its various types. A simple application of the gradient descent may not work well in a deep network because it is confronted with problems by going and returning around the local optimum. This situation is fixed by the momentum parameter which helps to update the gradient descent, which is necessary to reach an optimal point.

One of the crucial parts of developing neural network architecture is the selection of hyperparameters. The hyperparameters are variables set to specific values before the training process. In [60], a list of the most effective hyperparameters adopted by many researchers for model performance has been proposed. This list includes initial learning rate, mini-batch size, number of training iterations, momentum, number of hidden units, weight initialization, weight decay, regularization strength and more hyperparameters.

### 3.3 Some Known CNN Architectures

LeNet [12], AlexNet [16], VGGNet [33], GoogleNet [61] and ResNet [4] are among the best- known CNN architectures. AlexNet and VGG16, 16-layer (convolutional+fully connected layers) version of VGG-Net, form the basis for the FCN architectures to be addressed in the next section.

Developed by Alex Krizhevsky, Ilya Sutskever, and Geoff Hinton, AlexNet is the first study to make CNN popular in computer vision [16]. AlexNet was presented at the ILSVRC in 2012, and was the first in the competition to perform significantly better than the second and third architectures Even though AlexNet has an architecture similar to LeNet, convolutional layers with deeper and more specific are stacked on top of each other. AlexNet has been trained on more than 1 million high-resolution images containing 1000 different classes.

Developed by Simonian and Zisserman, VGGNet has two versions, called VGG-16 and VGG-19. The VGG-16 architecture was the second in the ILSVRC in 2014, where GoogleNet was first. Recently, ResNet seems to be a very advanced CNN model, but VGG-16 is preferred because of its simple architecture. VGG-16 has also been trained on more than 1 million high-resolution images containing 1000 different classes like AlexNet.

## 4 Semantic Image Segmentation and Fully Convolutional Networks

Except where it is in the image, there are many situations that need to be learned For example, you should know the locations of the objects around autonomous vehicles so that you can move without hitting anywhere. The process of determining the locations of the objects in the image can be realized by detecting (getting into the bounding box) or segmenting. As mentioned before, an application of semantic image segmentation for the autonomous vehicles will be implemented in this paper.

Image segmentation is the operation of separating images into specific groups that show similarity and labeling each pixel in an image as belonging to a given semantic class. In order for the image segmentation process to be considered successful, it is expected that the objects are independent of the other components in the image, as well as the creating of regions with the same texture and color characteristics. As an example in Fig. 14, each pixel in a group corresponds to the object class as a whole. These classes may represent an urban scene that is important for autonomous vehicles; traffic signs, cars, pedestrians, street lamps or sidewalks.

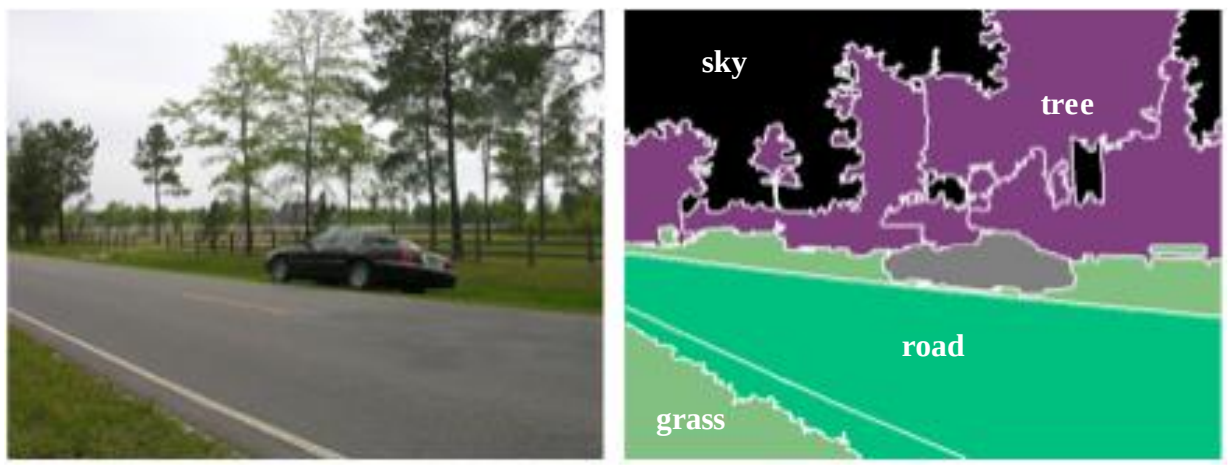

**Fig. 14.** A sample of semantic image segmentation [62]

As mentioned before, AlexNet and similar standard CNNs perform non-spatial prediction. For example, in the image classification, the network output is a single distribution of class probabilities. The CNN must be converted into the FCN in order to achieve a density prediction such as semantic image segmentation. Because, as described in [20], the fully connected layers of the CNNs give information about location. Therefore, in order to be able to carry out a semantic image segmentation, it is necessary to convolve the fully connected layers of the CNNs.

In the conversion to the FCN, the convolution part of the CNN can be completely reused. Fully convolutional versions of existing CNNs predict dense outputs with efficient inference and learning from arbitrary sized inputs. Both learning and inference present the whole visual in a single pass through forward computation and back-propagation, as shown in Fig. 15. The upsampling layers within the network allow learning on the network via pixel-level prediction and subsampling [20].

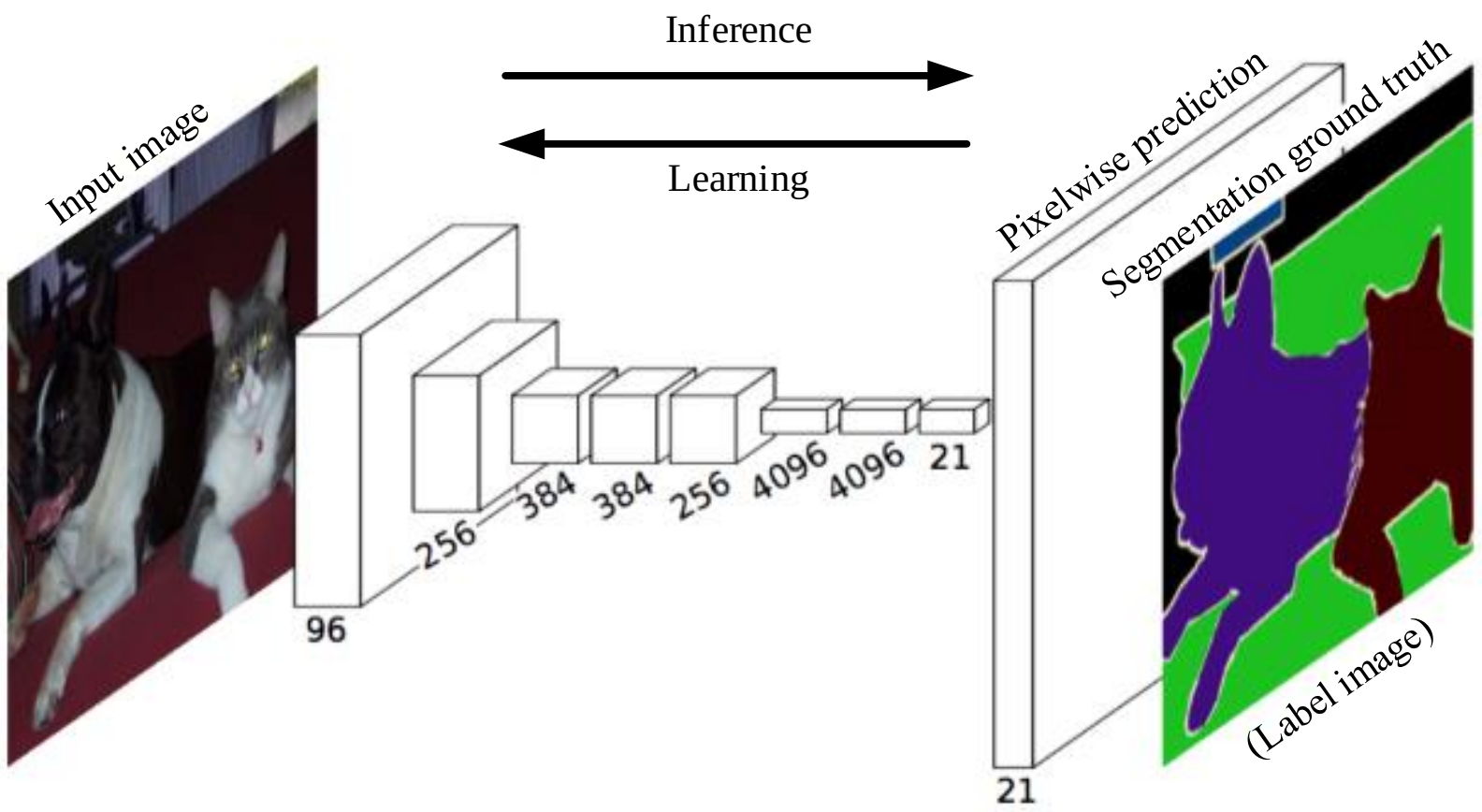

**Fig. 15.** Pixel-level prediction with FCN [20]

### 4.1 Semantic Image Segmentation

For semantic image segmentation application for autonomous vehicles, it may be thought that this can overcome its with the thing coming to mind classification network architectures such as AlexNet, GoogleNet and VGG-16. However, the models created by image classification architectures give the output of which class is the dominant object class in the image. That is, the output is a discrete probability distribution. If the image contains more than one object, the desired result cannot be obtained. Considering that classification models such as AlexNet are trained with a data set consisting of more than one million images of only one object in it, it is quite understandable. In addition, location information of the object in the image cannot be obtained with the classification networks. The situation can be understood when they are thought to have never been trained for this aim. The semantic image segmentation eliminates some of these deficiencies. Instead of estimating a single probability distribution for an entire image, the image is divided into several blocks and each block is assigned its own probability distribution. Very commonly, images are divided into pixel-levels and each pixel is classified. For each pixel in the image, the network is trained to predict which class the pixel belongs to. This allows the network not only to identify several object classes in the image but also to determine the location of the objects.

The datasets used for semantic image segmentation are images that are to be segmented and usually consist of label images of the same size as these images. The label image shows the ground truth limits of the image. The shapes in the label images are coded with colors to represent the class of each object. In some of the datasets, especially the SYNTHIA-Rand-CVPR16 dataset used in this paper, the label images consist of 24-bit 3 channel RGB images. In this case, the pixels must be indexed by creating color maps with RGB channel conversion.

## 4.2 Conversion from CNN to FCN

Semantic image segmentation only adds a spatial dimension to the image classification problem. Thus, several minor adjustments are sufficient to convert a classification neural network into a semantic segmentation neural network. It is implemented on AlexNet and VGG-16 architectures using the techniques of conversion to FCN architecture [20], which is necessary for the conventional image segmentation. The conversion to FCN is achieved by converting the fully connected layers of the CNN to convolutional layers. This conversion is shown generally in Fig. 16.

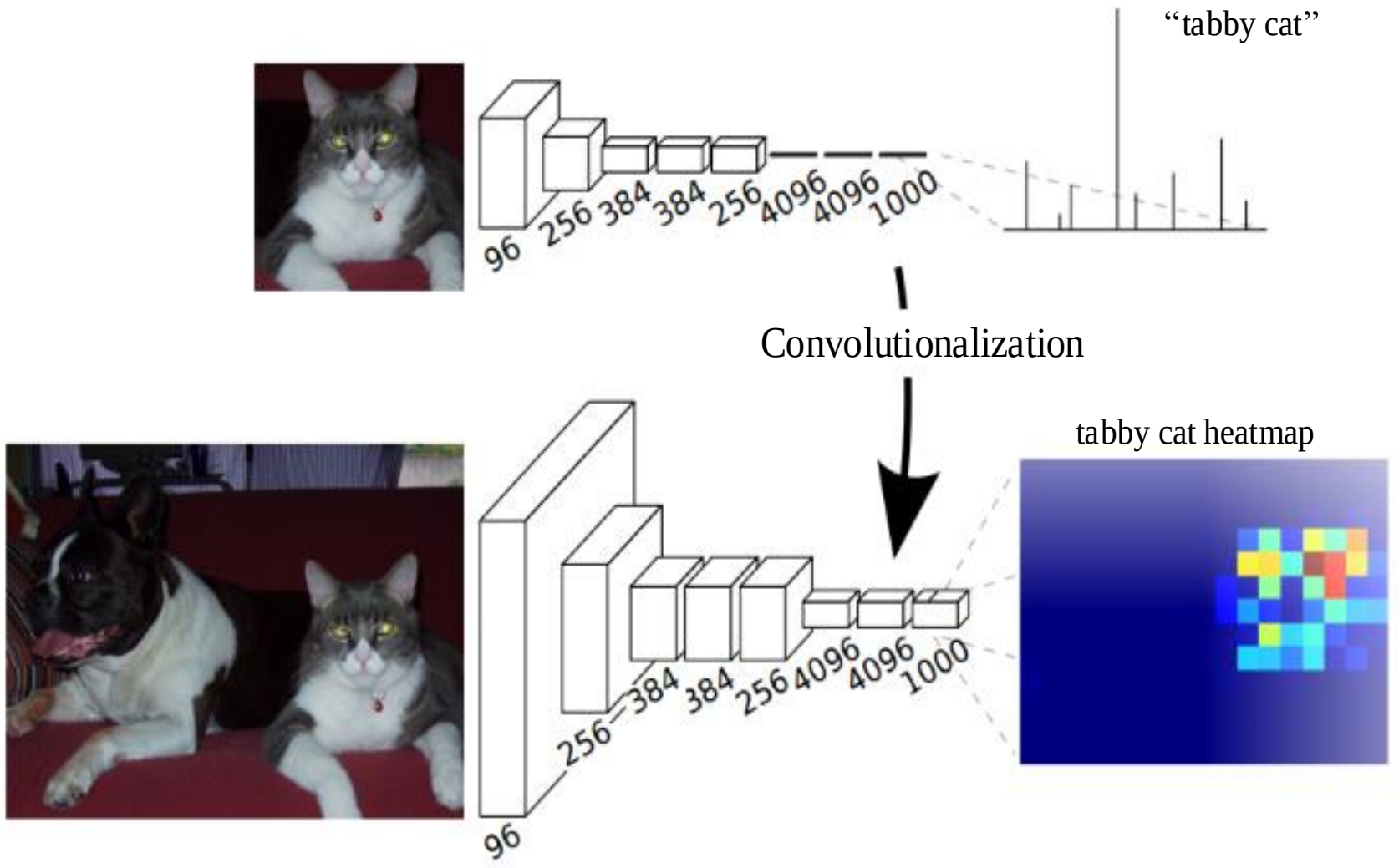

**Fig. 16.** Converting fully connected layers of CNN to convolutional layers [20]

In a fully connected layer, each output neuron calculates the weighted sum of the values in input, while in a convolutional layer, each filter calculates the weighted sum of the values in the receptive field. Although these operations seem to be exactly the same thing, they are the same only when the layer input has the same size as the receptive field. If the input is larger than the receptive field, then the convolutional layer slide input window and calculate another weighted sum. This repeats until the input image is scanned from left to right, from top to bottom. A fully connected layer must be replaced with a corresponding convolutional layer, the size of the filters must be set to the input size of the layer, and as many filters as the neurons in the fully connected layer must be used.

All connected layers in the AlexNet architecture can be converted to corresponding convolutional layers to obtain FCN architecture. This FCN has the same number of learned parameters as the basic CNN and the same computational complexity. Convolutionalization of a basic CNN brings considerable flexibility to the conversion to FCN. The FCN model is no longer limited to work with fixed input size 224×224, as in AlexNet. The FCN can process large images by scanning throughly, such as a sliding window, and the model generates one per 224×224 window rather than generating a single probability distribution for the entire input. Thus, the output of the network has become a tensor in the form of N×H×W.
where; N is the number of classes, H is the number of sliding windows (filters) along the vertical axis, and W is the the number of sliding windows along the horizontal axis.

In summary, the first significant step in the design of the FCN is completed by adding two spatial dimensions to the exit of the classification network.

The window number depends on the size of the input image, the size of the window, and the stride parameters used between the windows when the input image is scanned, as will be understood during the design of a FCN that generates a class probability distribution per window. Ideally, a semantic image segmentation model should generate a probability distribution per pixel in the image. When the input image passes through the sequential layers of convolutionalized AlexNet, coarse features are extracted. The purpose of semantic image segmentation is to interpolate for these coarse features to reconstruct a fine-tuned classification for each pixel in the input. This can easily be done with deconvolutional layers. The deconvolutional layers perform the inverse operation of their convolutional counterparts. Considering the output of the convolutional layer, the deconvolutional layer finds the input generating the output. As it can be remembered, the stride parameter in the convolutional or pooling layer is a measure of how much the window is to be slid when the input is processed, and how the output is subsampled accordingly. In contrast, the stride parameter in the deconvolutional layer is a measure of how the output is upsampled. The output volume of the deconvolutional layer, $D_o \times H_o \times W_o$, is calculated using the equations (13), (14) and (15).

$$D_o = N \tag{13}$$

$$H_o = s(H_i - 1) + f - 2p \tag{14}$$

$$W_o = s(W_i - 1) + f - 2p \tag{15}$$

where; s stride, p padding, f filter size, $H_i$ and $W_i$ input sizes, and N is the number of channels.

It is important to know how much of the activation of the last convolutional layer in the FCN architecture must be upsampled to obtain an output of the same size as the input image. The upsampling layer added to create FCN-AlexNet is shown to increase the output of the previous convolutional layer by 32 times. This means that in practice, the network has made a single prediction per 32×32 pixel block. This causes the contours of objects in the image to be segmented as rough. Fig. 17 shows FCN-AlexNet architecture.

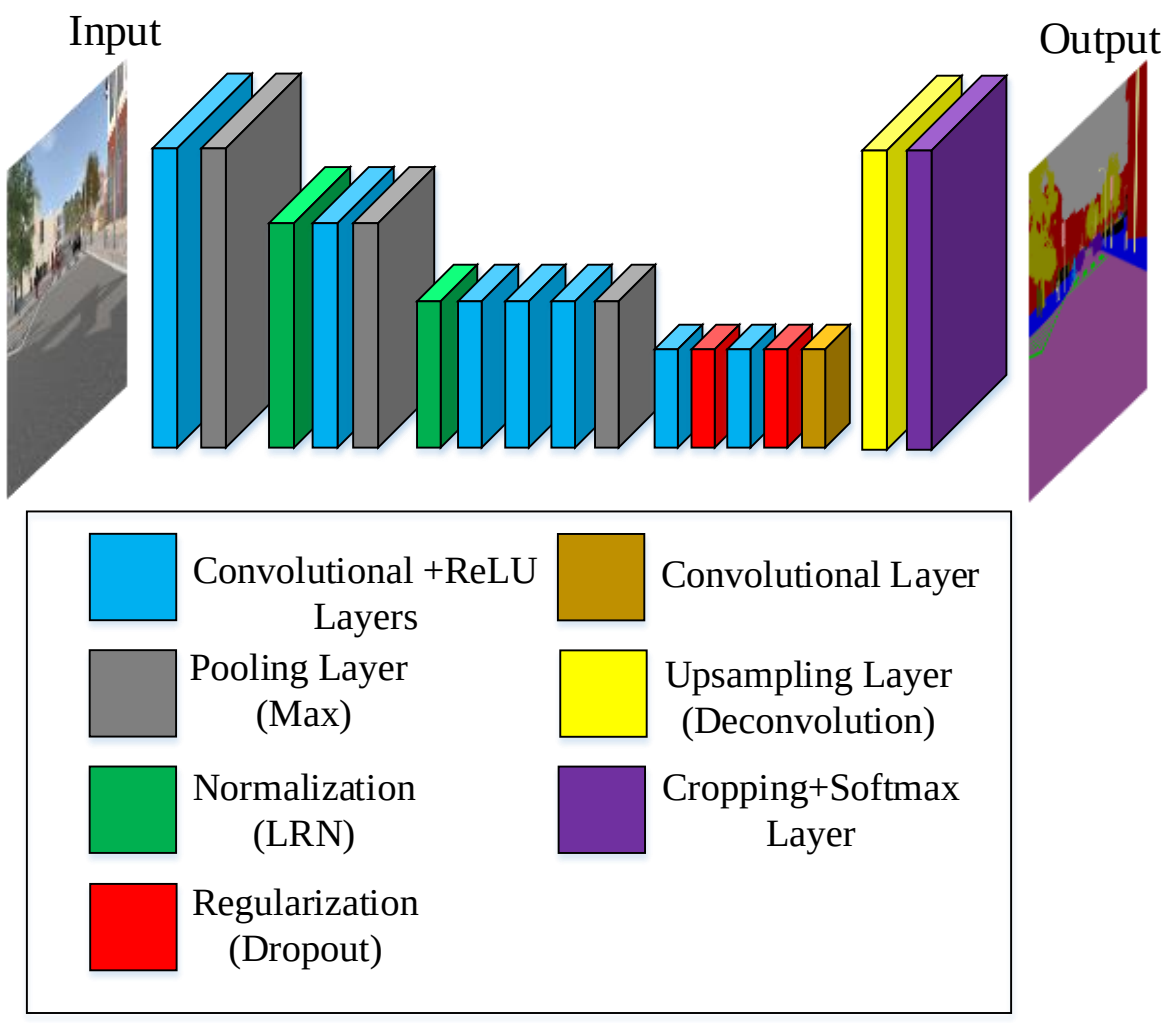

**Fig. 17.** FCN-AlexNet architecture

The article in [20] presents the idea of skip architecture for this restriction. The skip connections in this architecture have been added to redirect the outputs of the pooling layers pool3 and pool4 of the FCN architecture derived from VGG-16 directly to the network as shown in Fig. 18. These pooling layers work on low-level features and can capture more fine details.

The FCN architectures proposed in [20] are called FCN-8s, FCN-16s and FCN-32 according to the application of skip connections, converted into corresponding convolutional layers of fully connected layers in VGG-16. The visualization of these architectures is shown in Fig. 18.

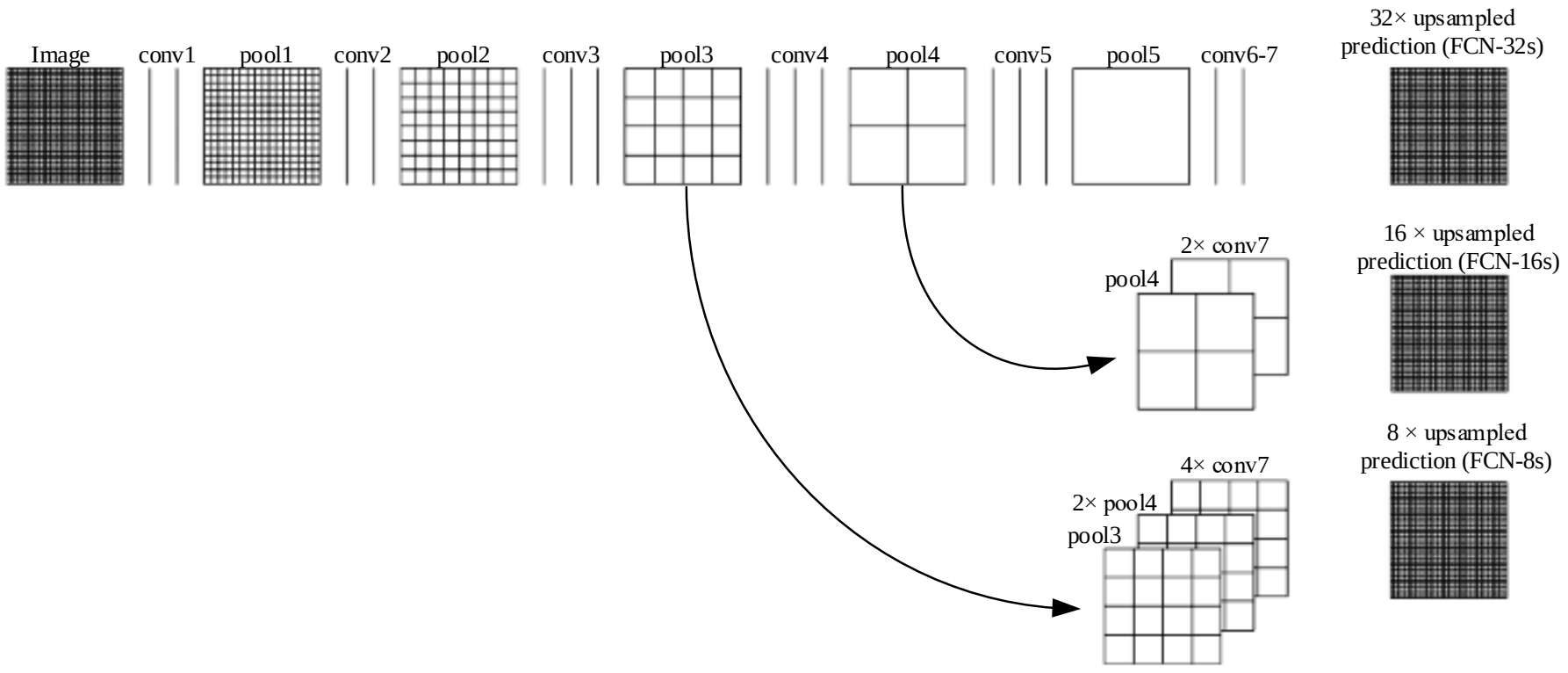

**Fig. 18.** Visualization of FCN-32s, FCN-16s and FCN-8s architectures [20]

In Fig. 18, the pooling layers are shown as grids expressing the spatial density, while the convolutional layers are shown as vertical lines. As it can be seen, the predictions in FCN-32s is upsampled at stride 32 to pixels in a single step without skip connection. The predictions in FCN-16s is combined at stride 16 from both *conv7* and *pool4* layers, allowing to predict finer details when detecting high-level semantic information. FCN-8s provides sharper predictions at stride 8 by adding the *pool3* layer. Thus, FCN-8s architecture allows to make fine-tuned predictions up to 8×8 pixel blocks.

# 5 Experimental Studies

In this paper, a semantic image segmentation application, which is useful for autonomous vehicles, was performed to observe the performance of the FCNs in semantic image segmentation. Four different popular FCN architectures were used separately for the application: FCN-AlexNet, FCN-8s, FCN-16s and FCN-32s.

The applications were implemented using Caffe framework in DIGITS platform on SYNTHIA-Rand-CVPR16 dataset and the segmentation performances of the used FCN architectures for experimental studies were compared. The studies were carried out on a desktop computer with 4th Generation Intel® Core i5 3.4 GHz processor, 8 GB RAM and NVIDIA GTX Titan X Pascal 12 GB GDDR5X graphics card. Thanks to the CUDA support of the graphics card, the GPU-based parallel computing power has been utilized in the computations required for the application.

## 5.1 SYNTHIA-Rand-CVPR16 Dataset

The SYNTHIA-Rand-CVPR16 dataset [40] has been generated to support semantic image segmentation in autonomous driving applications. The images in this dataset were created by portraying a virtual city with the Unity development platform [63]. The virtual environment allows them to freely place the desired components in the scene image and generate its semantic annotations without additional effort.

The SYNTHIA-Rand-CVPR16 dataset consists of a 13407 RGB image with a resolution of 960×720 taken from a virtual camera array randomly moving through the city, limited to the range [1.5m, 2m] from the ground. It also consists of ground truth images of the same size as these RGB images. The dataset images, which are taken under different conditions such as night and day, includes 12 object classes: sky, building, road, sidewalk, fence, vegetation, pole, car, sign, pedestrian, cyclist and void.

Sample images from the SYNTHIA-Rand-CVPR16 dataset are shown in Fig. 19. For the paper, a total of 13407 images of the dataset are used to train the network model, and the remaining 2681 (about 20% of the dataset) are used to validate the model.

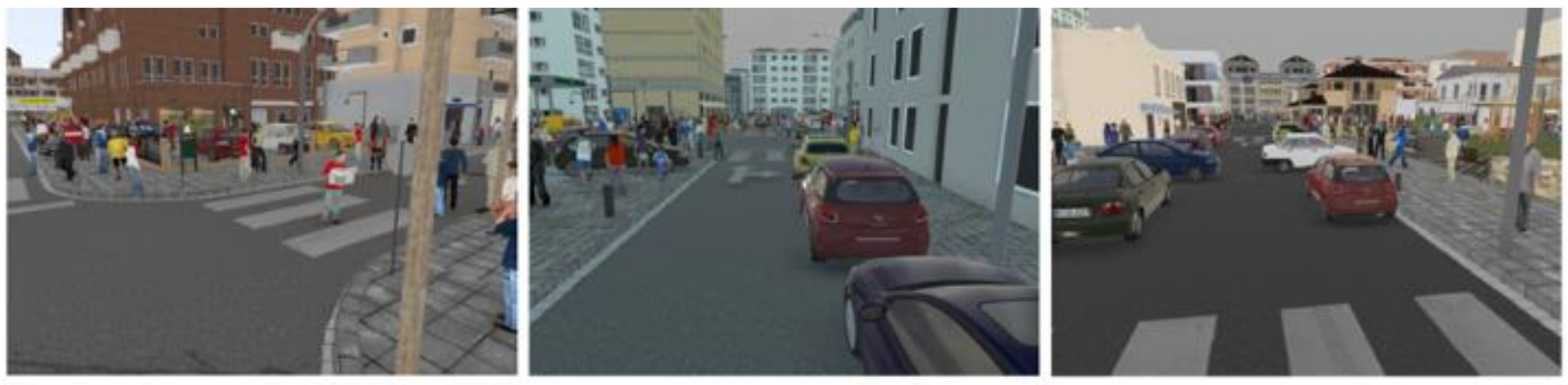

(a)

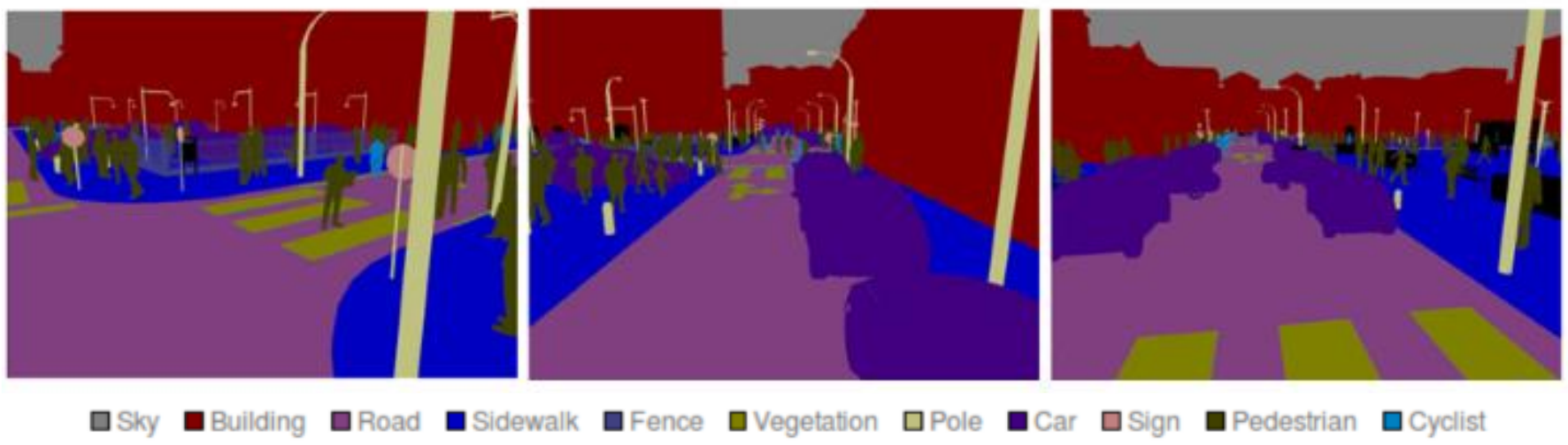

(b)

**Fig. 19.** Samples from the SYNTHIA-Rand-CVPR16 dataset: (a) Sample images, (b) Ground truth images with semantic labels

## 5.2 Training and Validations of the Models

The values of the network parameters determined for the application are given in Table 1. These parameters are generally set to be used for all the used FCN models in the application.

**Table 1.** The values of the network parameters

| Parameter | Value |
|---|---|
| Base learning rate | 0.0001 |
| Momentum | 0.9 |
| Weight decay | $10^{-6}$ |
| Batch size | 1 |
| Gamma | 0.1 |
| Maximum iteration | 321780 |

Stochastic Gradient Descent (SGD) as solver type and GPU as solver mode are selected. Epoch is set to 30. An epoch is a single pass through the full training set. Thus, for 10726 training images, 1 epoch is completed in 10726 iterations with 1 batch size and it is seen that the number of maximum iteration number for 30 epochs is 321780, as indicated in Table 1.

Initially, FCN-AlexNet model is trained using random weight initialization in DIGITS and the results in Fig. 20 are obtained.

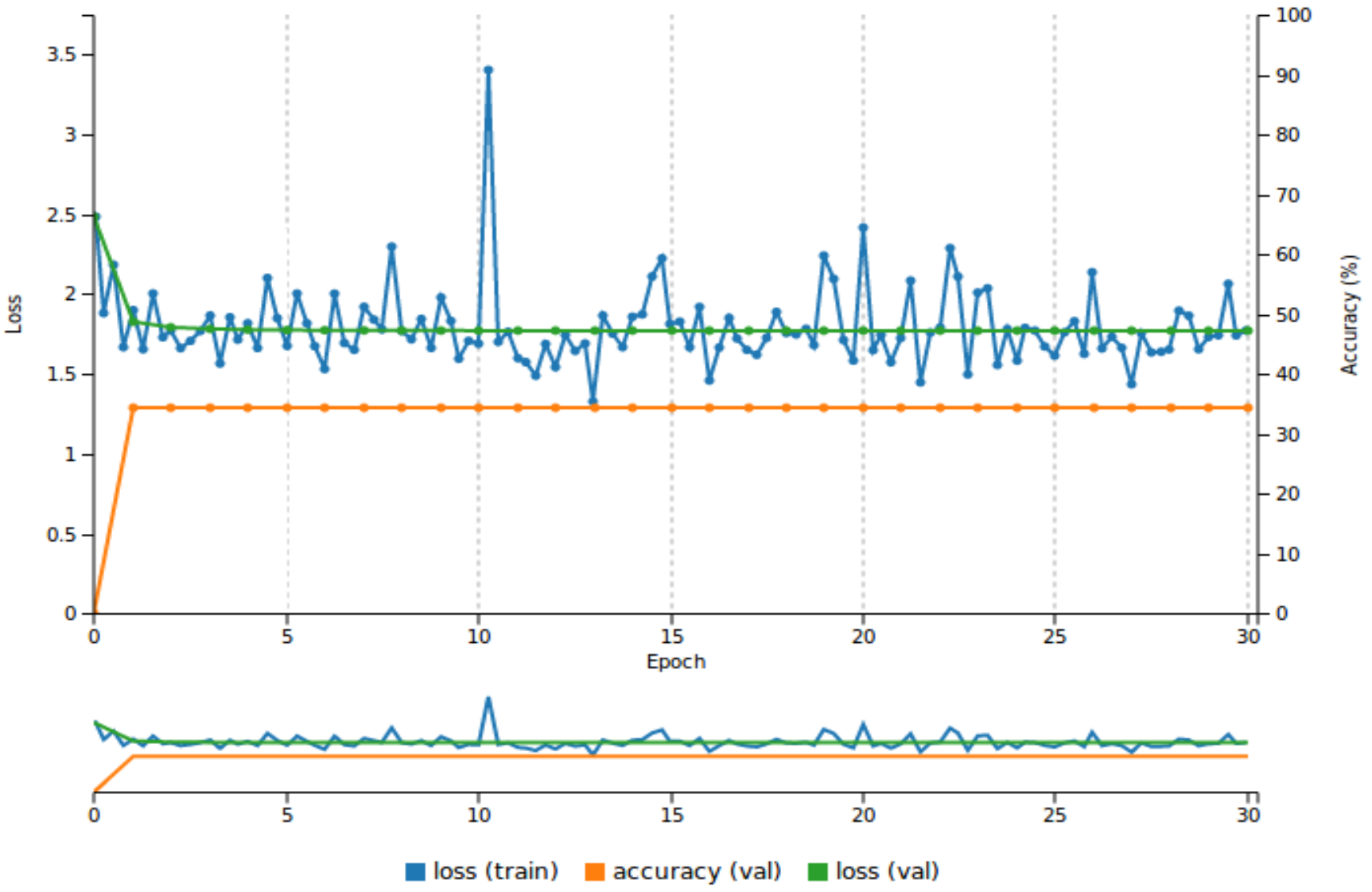

**Fig. 20.** Training/validation loss and validation accuracy when training FCN-AlexNet using random weight initialization

As shown in Fig. 20, performance is not satisfactory enough. Validation accuracy has reached a stationary point of about 35%. This means that only about 35% of the pixels in the validation set are correctly labeled. The training loss, which indicates that the network is not suitable for the training set, is parallel to the validation loss. When the trained model is tested on sample images in the validation set and visualized in DIGITS, it can be seen in Fig. 21 that the network classifies indiscriminately everything as building.

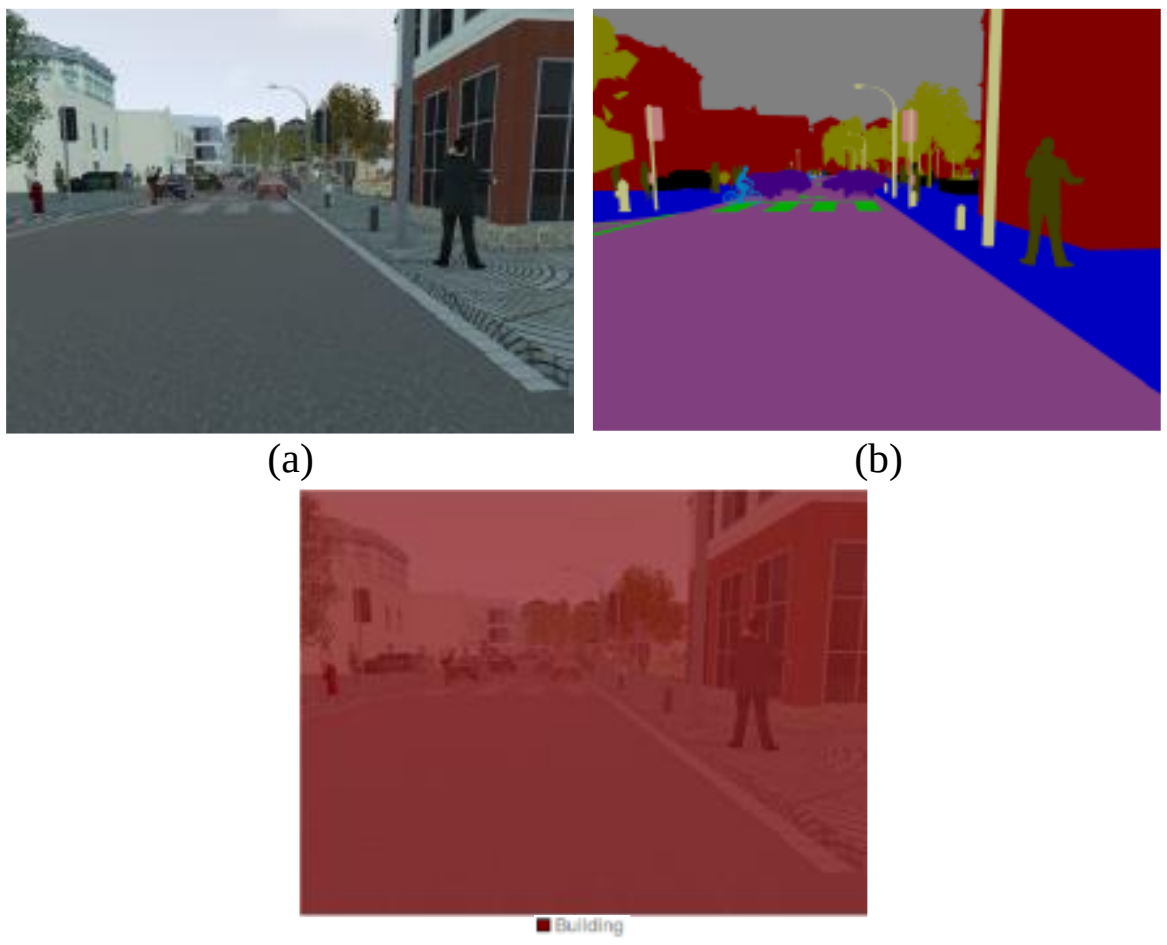

(a) (b) (c)

**Fig. 21.** A sample visualization of semantic image segmentation in DIGITS with FCN-AlexNet trained using random weight initialization: (a) Sample image, (b) Ground truth image, (c) Inference

With the Fig. 21, it is understood that the building is the most representative object class in the SYNTHIA-Rand-CVPR16 dataset, and that the network has learned to achieve approximately 35% accuracy by labeling everything as building.

There are several commonly accepted ways to improve a network that is not suited to the training set [64]. These:

- Increasing the learning rate, and reducing the batch size,
- Increasing the size of the network model,
- Transfer learning.

Information learned by a deep network can be used to improve the performance of another network and this process is very successful for computer vision applications. For this reason, while learning the required models for the application, transfer learning was used.

Recent developments in machine learning and computer vision are first achieved through the use of common criteria. It does not have to start from randomly initialized weights to train a model. Transfer learning is a reuse of information that a network learns in another dataset to improve the performance of another network [65]. A network is trained on any data and gains knowledge from this data, compiled as weights of the network. These weights can be transferred to any network. In other words, instead of training the network from scratch, learned features can be transferred to the network.

Transfer learning is often preferred in the computer vision field, since many low-level features such as line, corner, shape, and texture can be immediately applied to any dataset via CNNs.

Models trained and tested on high-variance standard datasets usually owe their successes to strong features [65]. Transfer learning allows to use a model that learns fairly generalized weights trained on a large dataset such as ImageNet and allows fine-tuning to adapt the situation of the network to be used.

It is very logical to transfer learning from image classification dataset such as ImageNet since the image segmentation has a classification at the pixel level. This process is quite easy using Caffe. However, Caffe cannot automatically carry the weights from AlexNet to FCN-AlexNet because AlexNet and FCN-AlexNet have different weight formats. Moving these weights can be done using the Python command line "net_surgery.py" in DIGITS repository in Github. The function of net_surgery.py is to transfer weights from fully connected layers to convolutional equivalents [64].

Also, another possible problem is how to start the upsampling layer added to create FCN-AlexNet since the upsampling layer is not part of the original AlexNet model. In [20], it is recommended that the corresponding weights are first randomly initiated and the network learns them. Later, however, the authors realized that it is easy to initialize these weights by doing bilinear interpolation, the way that the layer just acts like a magnifying glass [64].

As previously mentioned, training of FCN-AlexNet model was performed using the pre-trained model obtained by adapting the AlexNet model trained on the ImageNet dataset and the results in Fig. 22 were obtained.

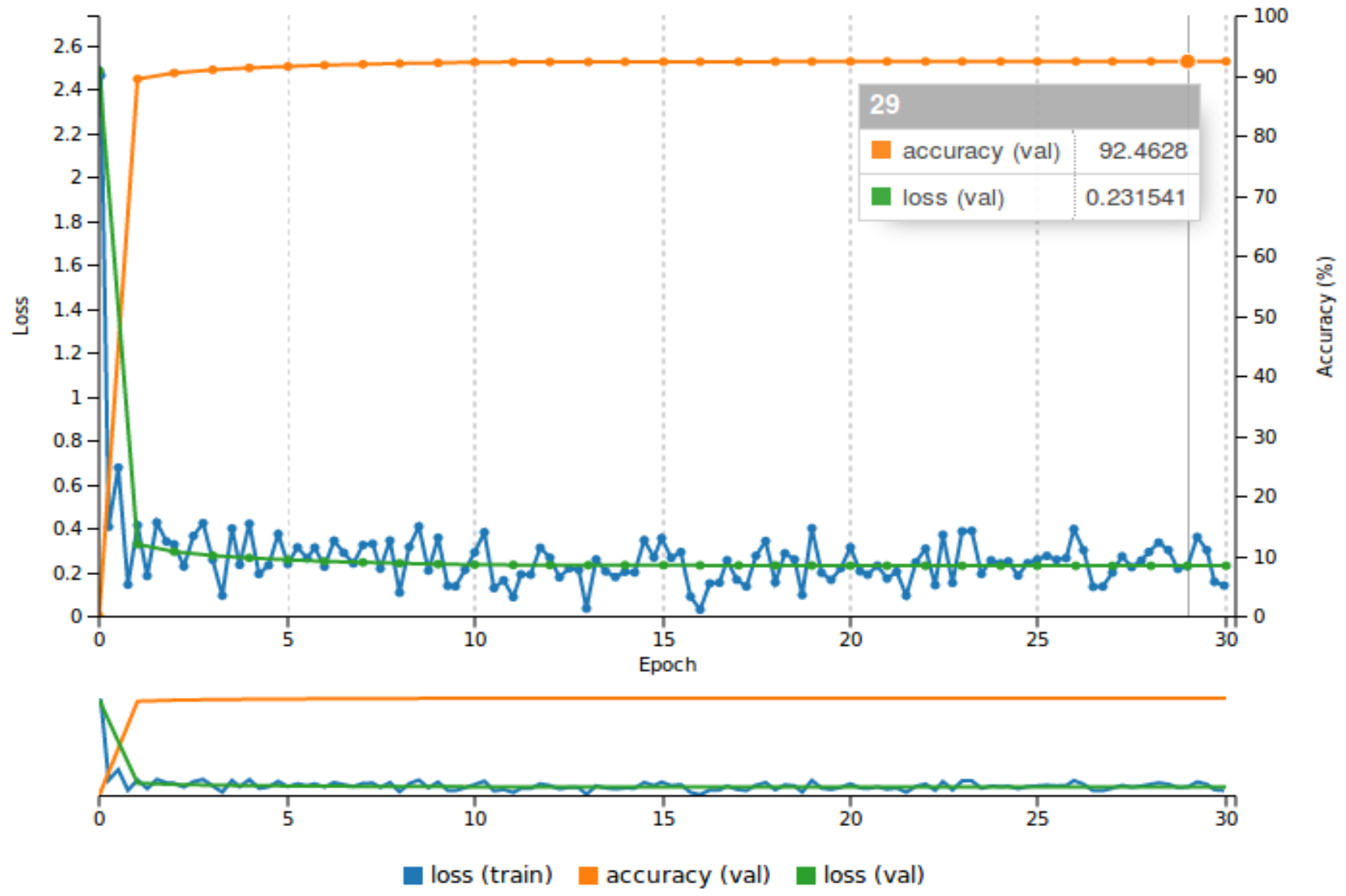

**Fig. 22.** Training/validation loss and validation accuracy when training FCN-AlexNet using pre-trained model

Fig. 22 shows that using the pre-trained FCN-AlexNet model, the validation accuracy quickly exceeded 90%, and the model achieved the highest accuracy at 92.4628% in 29th epoch. This means that 92.4628% of the pixels in the validation set of the model obtained in 29th epoch are labeled correctly. It has been shown to have fairly high accuracy compared to FCN-AlexNet initialized randomly weights.

When tested for sample images using the model obtained in 29th epoch, a semantic image segmentation was performed many times more satisfactorily by detecting different object classes as shown in Fig. 23 and Fig. 24. However, it can be clearly seen that the object contours are very rough.

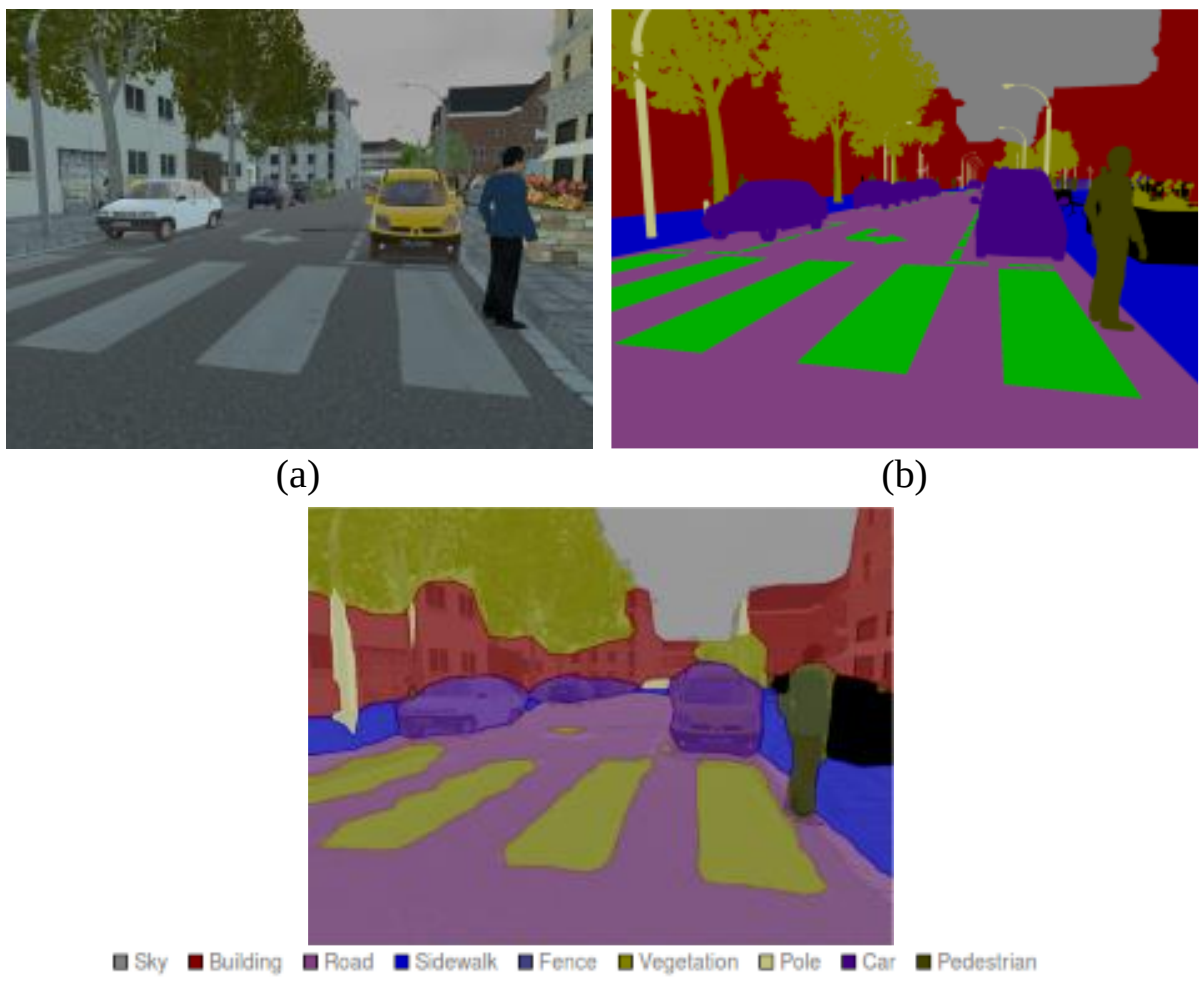

(c)

**Fig. 23.** A sample visualization of semantic image segmentation in DIGITS with FCN-AlexNet trained using pre-trained model-1: (a) Sample image, (b) Ground truth image, (c) Inference

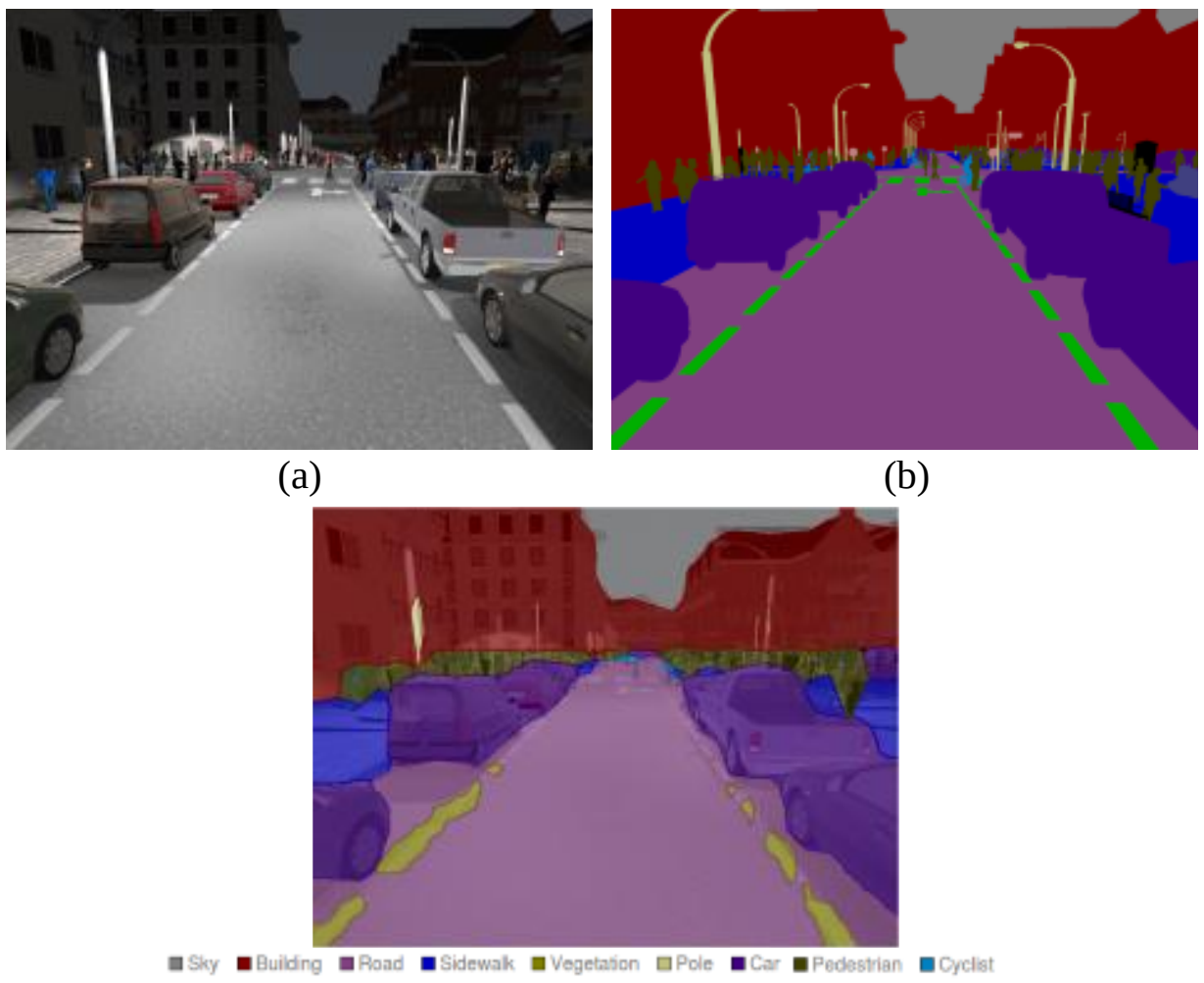

(c)

**Fig. 24**. A sample visualization of semantic image segmentation in DIGITS with FCN-AlexNet trained using pre-trained model-2: (a) Sample image, (b) Ground truth image, (c) Inference

FCN-8s network is used to further improve the precision and accuracy of the segmentation model. Using the pre-trained model in the PASCAL VOC dataset, validation accuracy of FCN-8s quickly exceeded 94% as shown in Fig. 25. The model reached to the highest accuracy with 96% in 30th epoch.

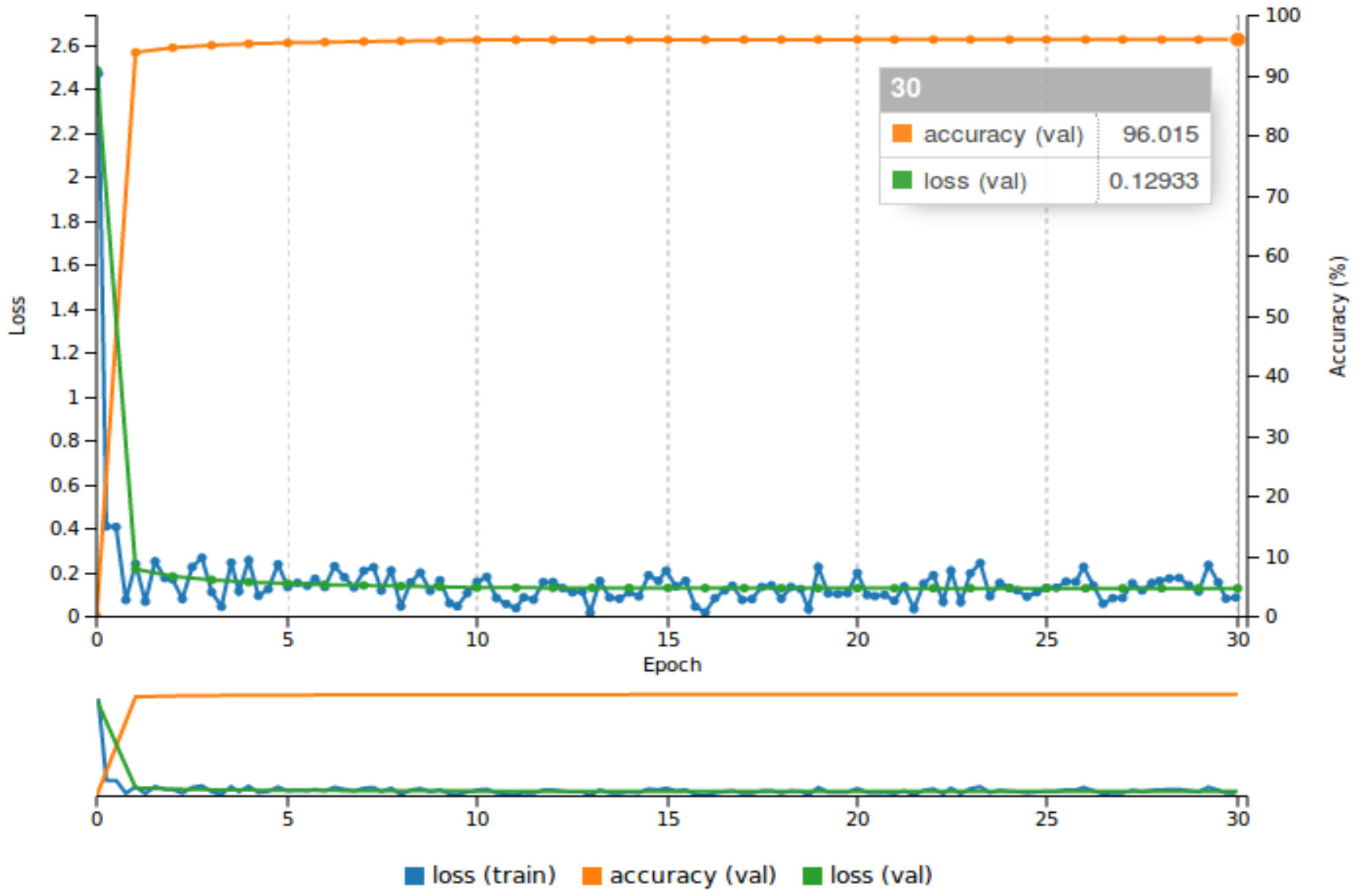

**Fig. 25**. Training/validation loss and validation accuracy when training FCN-8s using pre-trained model

More importantly, when tested for sample images using the model obtained in 30th epoch, much sharper object contours are shown, as shown in Fig. 26 and Fig. 27.

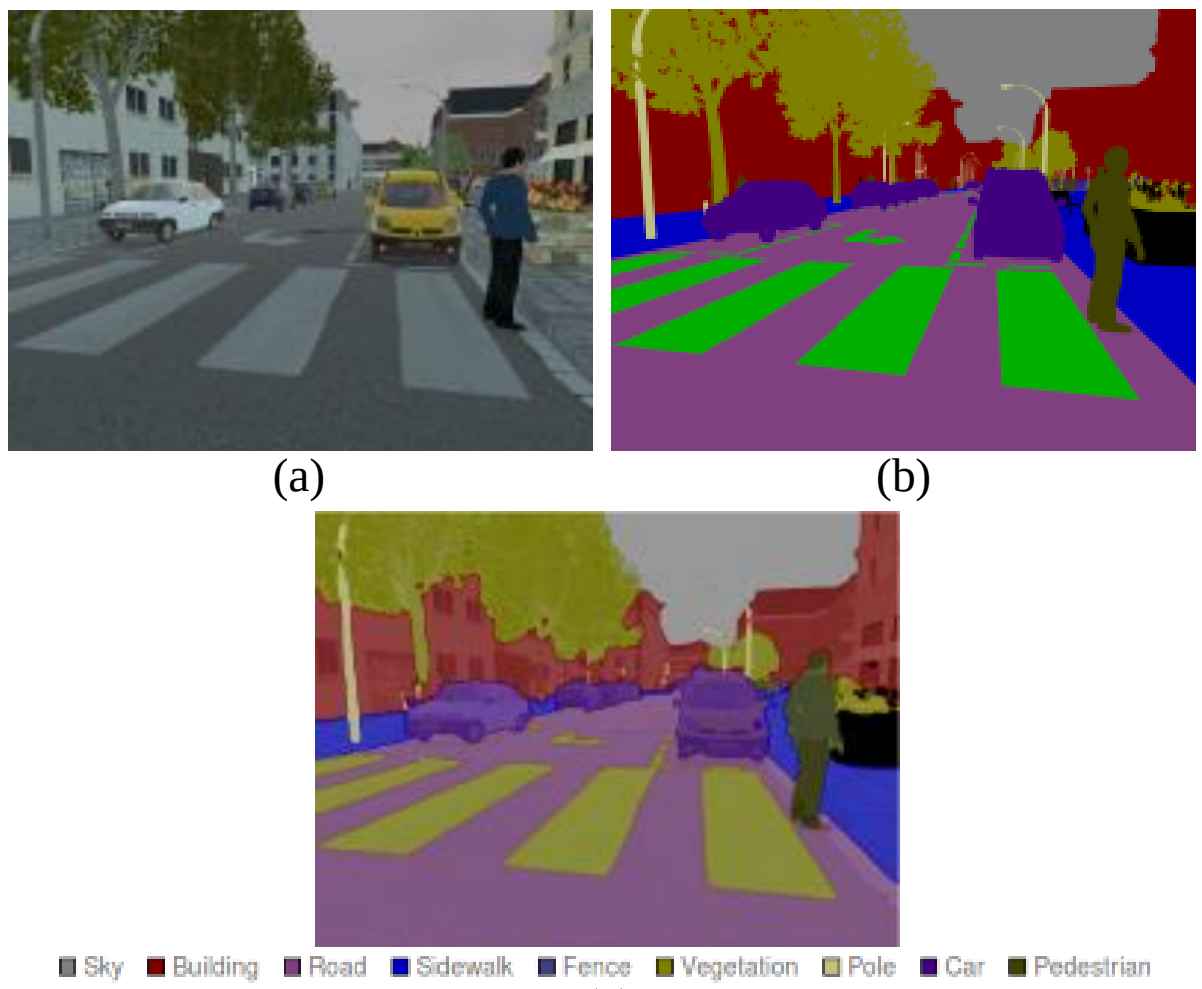

(a) (b) (c)

**Fig. 26.** A sample visualization of semantic image segmentation in DIGITS with FCN-8s trained using pre-trained model-1: (a) Sample image, (b) Ground truth image, (c) Inference

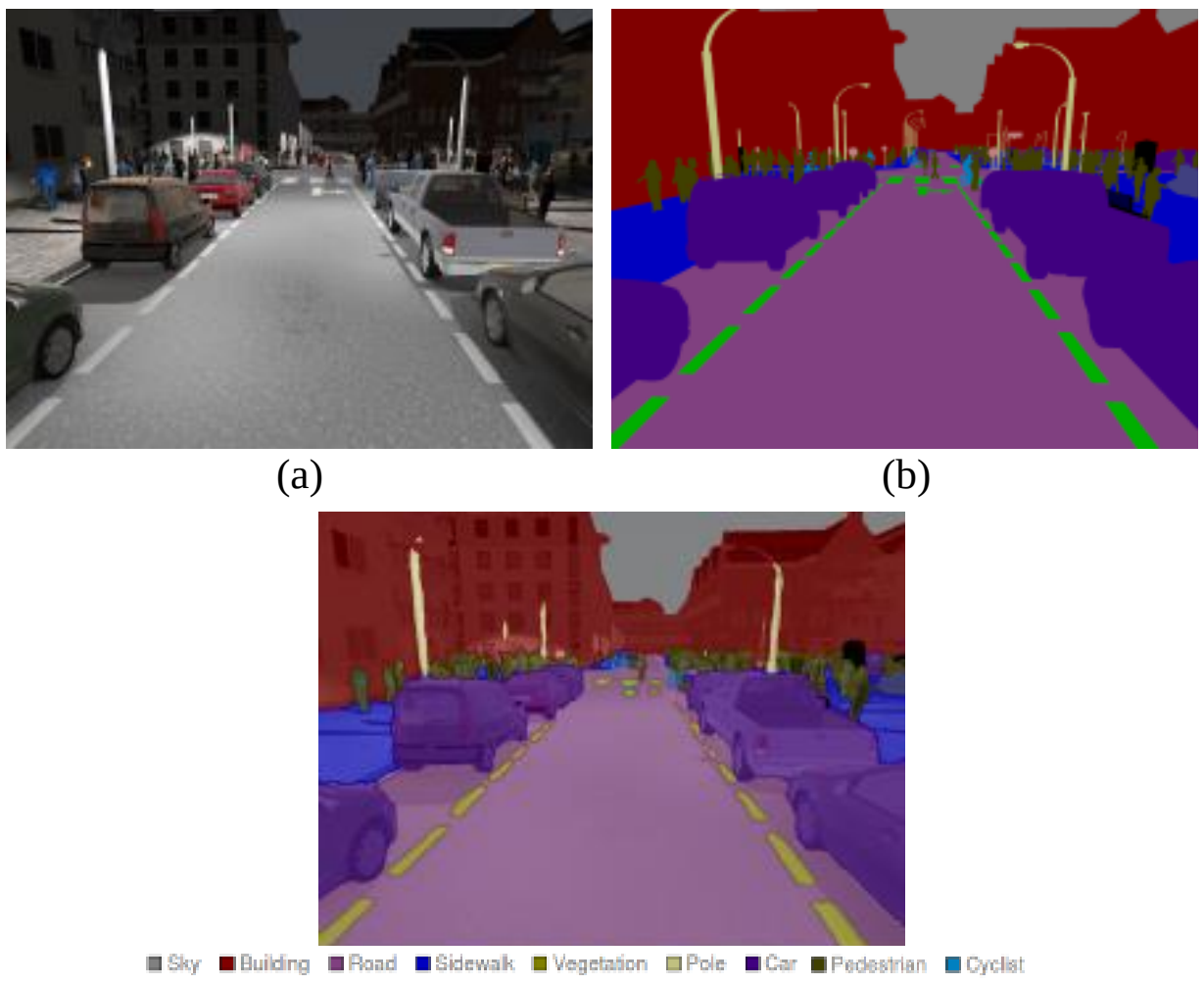

(c)

**Fig. 27.** A sample visualization of semantic image segmentation in DIGITS with FCN-8s trained using pre-trained model-2 (a) Sample image, (b) Ground truth image, (c) Inference

FCN-8s architecture has been shown to provide segmentation with sharper object contours than FCN-AlexNet, which makes predictions in 32×32 pixel blocks, as it can make predictions at a fine-tuning down to 8×8 pixel blocks. Similarly, trainings of the models have been carried out by FCN-16s and FCN-32s architectures and it can be seen in Fig. 28 and Fig. 29 that the validation accuracy has exceeded 94% rapidly in a similar manner to FCN-8s. The highest validation accuracy was reached in 30th epoch as in FCN-8s with 95.4111% and 94.2595% respectively. Besides, Fig. 30 shows the comparison of segmentation inferences on the same images selected using FCN-AlexNet, FCN-8s, FCN-16s and FCN-32s architectures.

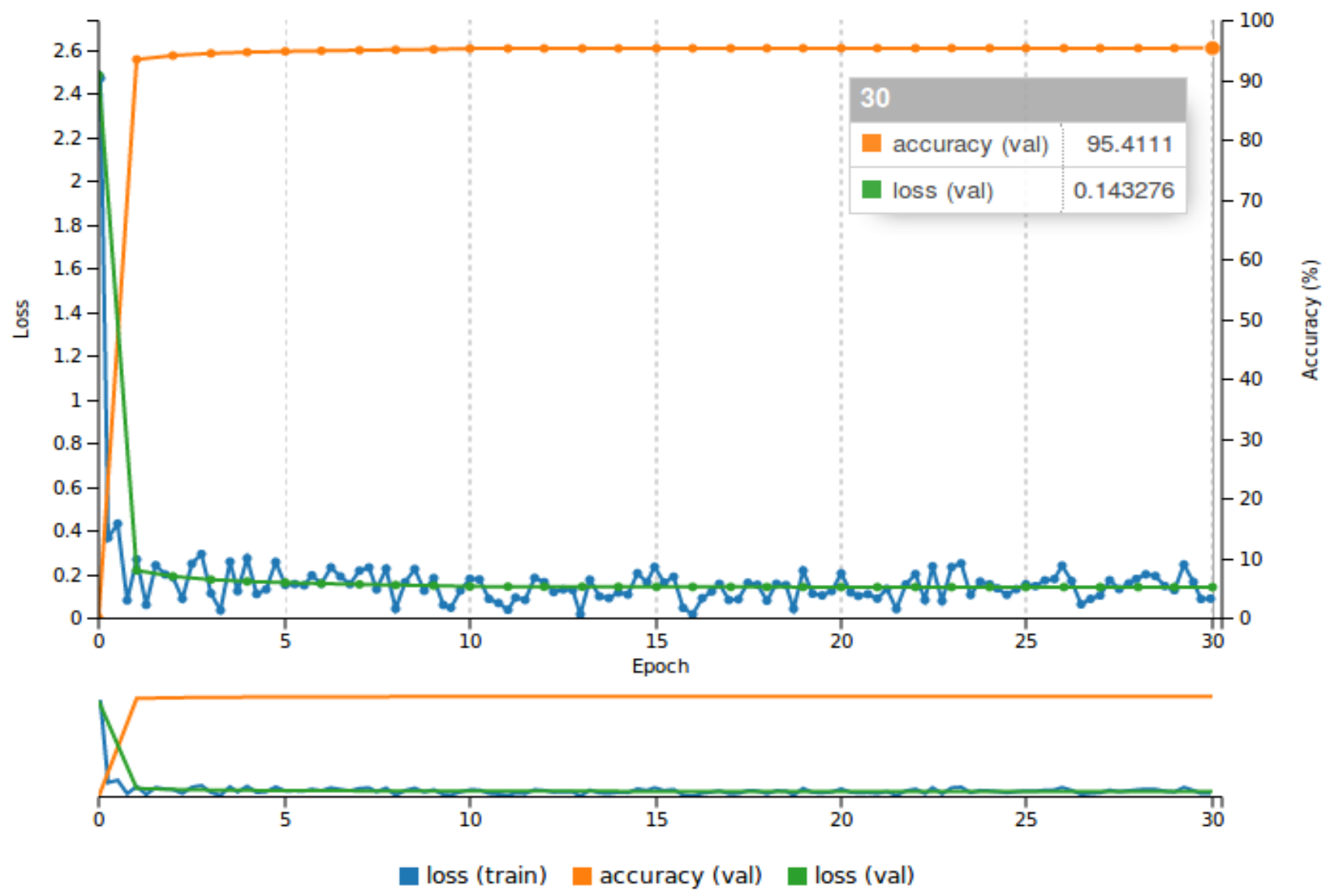

**Fig. 28.** Training/validation loss and validation accuracy when training FCN-16s using pre-trained model

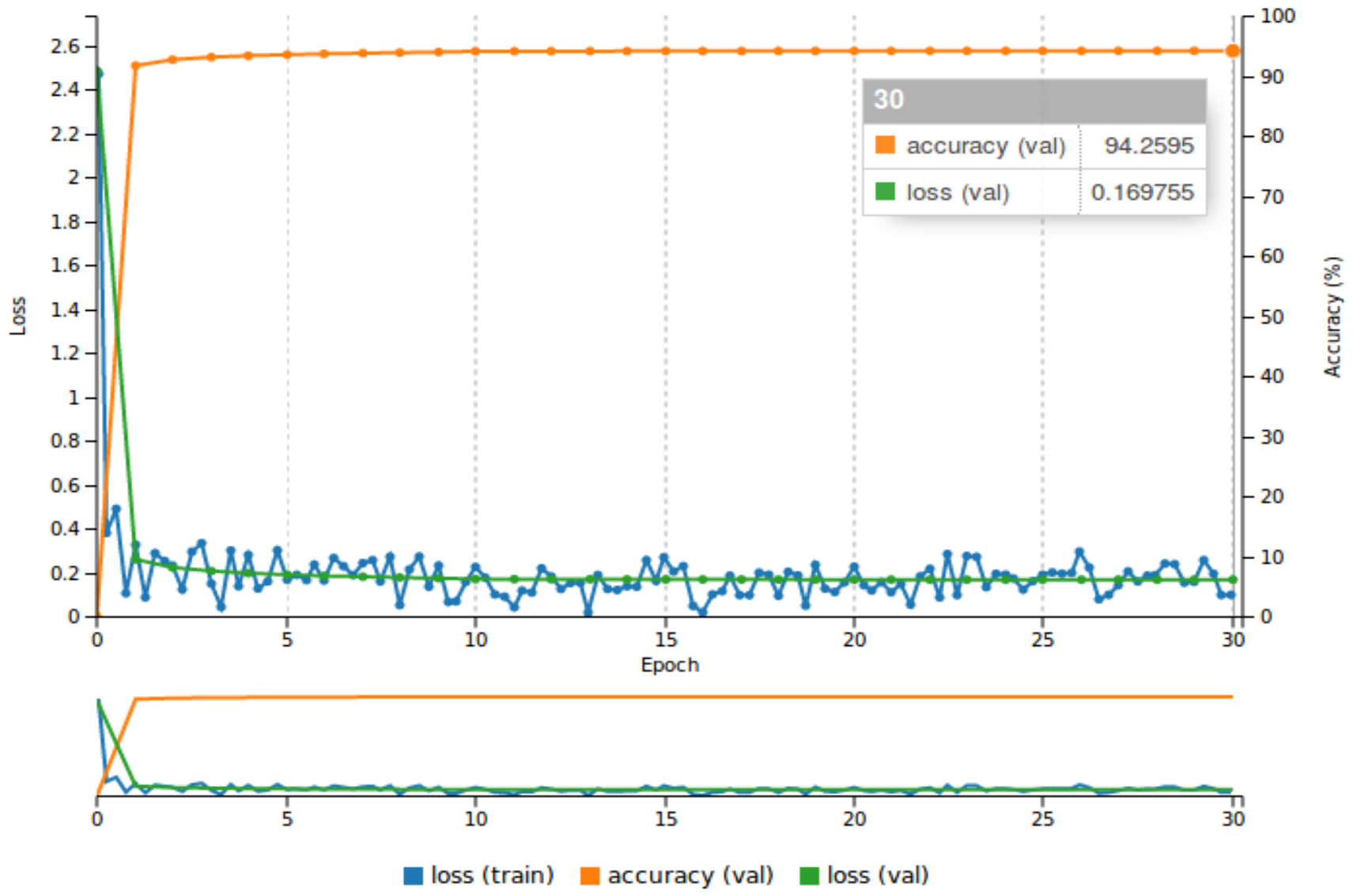

**Fig. 29.** Training/validation loss and validation accuracy when training FCN-32s using pre-trained model

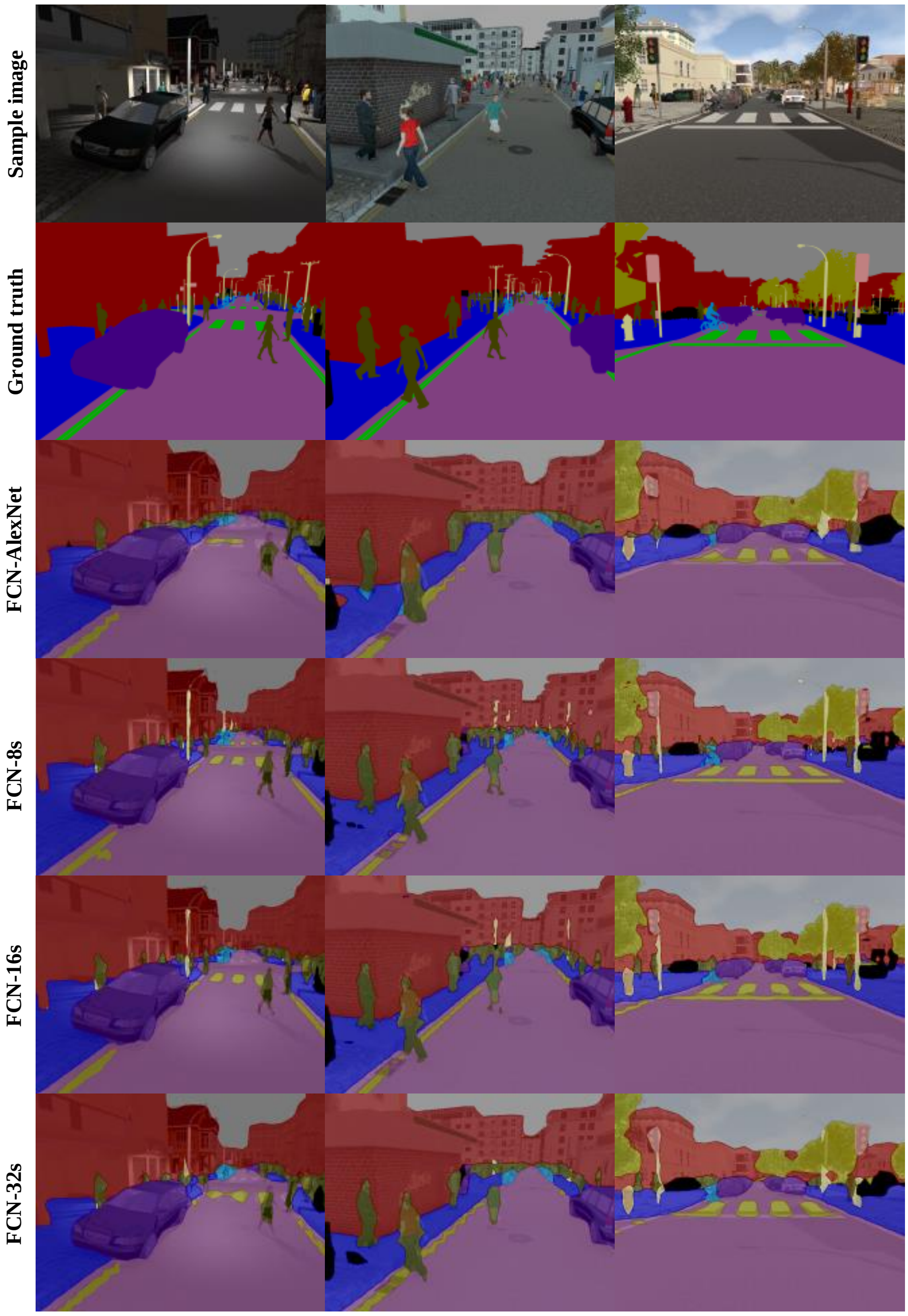

**Fig. 30.** Comparison of segmentation inferences according to the used FCN architectures for sample images

When the segmentation process is analyzed considering the sample and the ground truth images in Fig. 30, it has been seen that the object contours are roughly segmented in the segmentation process performed with FCN-AlexNet model. Moreover, the fact that fine details such as pole could not be made out and segmented showed another limitation of this model for the application. With FCN-8s model, contrary to FCN-AlexNet, object contours are segmented sharply and the segmentation inferences are more similar to the ground truth images. Furthermore, the fact that the object classes can be detected completely indicates that FCN-8s is useful. Although FCN-16s model is not as sharp as FCN-8s, it can be seen that the object contours can be segmented successfully. Finally, when the segmentation inferences of FCN-32s model are analyzed, it can be said that the segmentations very close to FCN-AlexNet have been realized but may be a more useful model with small differences.

The training times of the trained models for semantic image segmentation in this paper are given in Fig. 31.

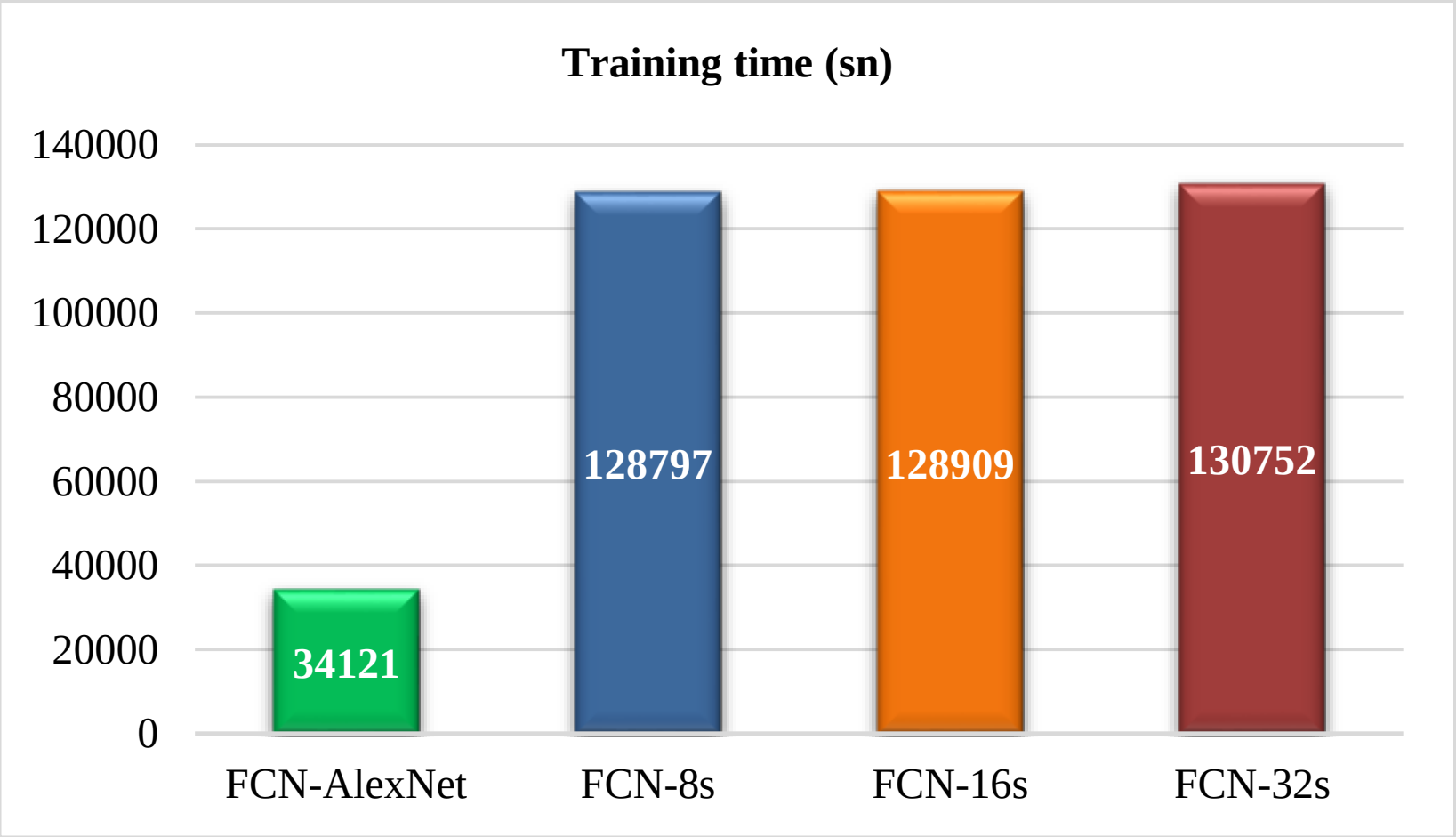

**Fig. 31.** Training times of the models

It was seen that the training time spent on FCN-AlexNet is considerably lower than the other FCN models with very close training times.

## 6 Conclusions

For the application, firstly, FCNs are trained separately and the validation accuracy of these trained network models is compared on the SYNTHIA-Rand-CVPR16 dataset. Approximately 80% of the images in the dataset are used for the training phase and the rest are used during validation phase to validate the validity of the models. In addition, image segmentation inferences in this paper are visualized to see how precisely the used FCN architectures can segment objects.

Maximum validation accuracies of 92.4628%, 96.015%, 95.4111% and 94.2595% are achieved with FCN-AlexNet, FCN-8s, FCN-16 and FCN-32s models trained using weights in pre-trained models, respectively. Although these models can be regard-ed as successful at first sight when the accuracies are over 90% for the four models, it is seen that the object contours are roughly segmented in the segmentation process performed with FCN-AlexNet model. The impossibility of segmenting some object classes with small pixel areas is another limitation of FCN-AlexNet model. The segmentation inferences of FCN-32s model are also very close to FCN-AlexNet, but with this model it is seen that some better results can be obtained. However, with FCN-8s model, object contours are sharply segmented and the segmentation inferences are more similar to the ground truth images. Although FCN-16 models are not as sharp as FCN-8s, it is seen that the object contours are successfully segmented according to the others.

When training times of FCN models are compared, it is seen that the training time spent on FCN-AlexNet is about one-fourth of the other FCN models with very close training times. However, considering that training of the model is carried out once for the application, it can be said that it does not have a very important place in the choice of the appropriate model. Therefore, it can be easily stated that the most suitable model for the application is FCN-8s.

The obtained experimental results show that the FCNs from deep learning approaches are suitable for semantic image segmentation applications. In addition, it has been understood that the FCNs are network structures in models that address many pixel-level applications, especially semantic image segmentation.

**Acknowledgements.** We gratefully acknowledge the support of the NVIDIA Corporation, who donated the NVIDIA GTX Titan X Pascal GPU that was used for this research under the NVIDIA GPU Grant program.